\documentclass[letterpaper,twocolumn,9pt,accepted=2024-10-30]{quantumarticle}

\pdfoutput=1
\usepackage[utf8]{inputenc}
\usepackage[english]{babel}
\usepackage[T1]{fontenc}

\usepackage{amsmath,amsthm,amssymb}
\usepackage{graphicx}
\usepackage{color}
\usepackage{array}
\usepackage{braket}
\usepackage{bbm}
\usepackage[square,numbers]{natbib}

\usepackage[hidelinks]{hyperref}
\hypersetup{
	colorlinks=true
}
\usepackage{cleveref}

\usepackage{pifont}%
\usepackage{bbold}%
\usepackage{braket}
\usepackage[dvipsnames]{xcolor} %
\usepackage{tikz}
\usepackage{qcircuit}
\usepackage{amsmath, amssymb, graphicx, bm}
\usepackage{subcaption}
\usepackage{comment}
\usepackage{pifont}%
\newcommand{\cmark}{\ding{51}}%
\newcommand{\xmark}{\ding{55}}%

\definecolor{mtl_red}{HTML}{D51B30}
\definecolor{mtl_green}{HTML}{007241}
\definecolor{mtl_blue}{HTML}{1E4596}
\definecolor{mtl_purple}{HTML}{894997}
\definecolor{mtl_yellow}{HTML}{F4B344}

\DeclareMathOperator{\haf}{haf}
\DeclareMathOperator{\lhaf}{lhaf}
\DeclareMathOperator{\mtl}{mtl}

\DeclareMathOperator{\ham}{ham}

\DeclareMathOperator{\lmtl}{lmtl}
\DeclareMathOperator{\per}{per}
\DeclareMathOperator{\Tr}{Tr}
\DeclareMathOperator{\fdiag}{fdiag}

\newcommand{\rh}{\hat{\bm{r}}}

\newcommand{\qh}{\hat{q}}

\newcommand{\ph}{\hat{p}}

\newcommand{\z}{\bm{\zeta}}
\newcommand{\zh}{\hat{\bm{\zeta}}}
\newcommand{\zb}{\bar{\bm{\zeta}}}
\renewcommand{\a}{\alpha}
\newcommand{\ah}{\hat{a}}
\newcommand{\ab}{\bar{\alpha}}
\renewcommand{\c}{\alpha^{*}}
\newcommand{\ch}{\hat{a}^{\dag}}
\newcommand{\cb}{\bar{\alpha}^{*}}
\newcommand{\nh}{\hat{n}}
\newcommand{\nb}{\bar{n}}
\newcommand{\mb}{\bar{m}}

\newcommand{\R}{\bm{R}}
\renewcommand{\O}{\bm{\Omega}}
\newcommand{\Z}{\bm{Z}}
\renewcommand{\S}{\bm{\Sigma}}
\newcommand{\s}{\Sigma} %
\newcommand{\V}{\bm{V}}
\newcommand{\I}{\mathbb{1}}
\newcommand{\X}{\bm{X}}

\newcommand{\M}{\bm{M}}
\newcommand{\m}{M} %
\newcommand{\N}{\bm{N}}
\newcommand{\n}{N} %
\newcommand{\A}{\bm{A}}

\newcommand{\U}{\bm{U}}

\newcommand{\js}{\bm{j}}

\newcommand{\bs}{\pmb{\beta}}
\newcommand{\gs}{\pmb{\gamma}}

\renewcommand{\l}{_{\ell}}
\renewcommand{\d}{^{\dag}}

\newcommand\crule[3][black]{\textcolor{#1}{\rule{#2}{#3}}} %

\begin{document}
	\title{Photon-number moments and cumulants of Gaussian states} 
\author{Yanic Cardin}
\email{yanic.cardin@polymtl.ca}
\author{Nicol\'as Quesada}
\email{nicolas.quesada@polymtl.ca}
\affiliation{D\'epartement de g\'enie physique, \'Ecole Polytechnique de Montr\'eal, Montr\'eal, QC, H3T 1J4, Canada}
	\begin{abstract}
We develop closed-form expressions for the moments and cumulants of Gaussian states when measured in the photon-number basis. We express the photon-number moments of a Gaussian state in terms of the loop Hafnian, a function that when applied to a $(0,1)$-matrix representing the adjacency of a graph, counts the number of its perfect matchings. Similarly, we express the photon-number cumulants in terms of the Montrealer, a newly introduced matrix function that when applied to a $(0,1)$-matrix counts the number of Hamiltonian cycles of that graph. Based on these graph-theoretic connections, we show that the calculation of photon-number moments and cumulants are $\#P-$hard. Moreover, we provide an exponential time algorithm to calculate Montrealers (and thus cumulants), matching well-known results for Hafnians. We then demonstrate that when a uniformly lossy interferometer is fed in every input with identical single-mode Gaussian states with zero displacement, all the odd-order cumulants but the first one are zero. Finally, we employ the expressions we derive to study the distribution of cumulants up to the fourth order for different input states in a Gaussian boson sampling setup where $K$ identical states are fed into an $\ell$-mode interferometer. We analyze the dependence of the cumulants as a function of the type of input state, squeezed, lossy squeezed, squashed, or thermal, and as a function of the number of non-vacuum inputs. We find that thermal states perform much worse than other classical states, such as squashed states, at mimicking the photon-number cumulants of lossy or lossless squeezed states.
	\end{abstract}
	\maketitle

\section{Introduction}
Gaussian states, those that have a Gaussian Wigner function, are some of the most well studied continuous-variable systems~\cite{weedbrook2012gaussian,ferraro2005gaussian}. Their statistics under continuous outcome measurement (such as homodyne or heterodyne) are well understood~\cite{serafini2017quantum}. Moreover, they are some of the easiest non-classical states of light to prepare experimentally~\cite{breitenbach1997measurement,lvovsky2015squeezed,loudon1987squeezed}. While the statistics of Gaussian measurements on Gaussian states are well understood, this is not the case for non-Gaussian measurements such as photon-number or threshold measurements. Only recently have photon-number probabilities been studied~\cite{hamilton2017gaussian,kruse2019detailed,quesada2019franck}, motivated by the introduction of Gaussian Boson Sampling (GBS), a subuniversal model of quantum computation that can demonstrate quantum computational advantage~\cite{hamilton2017gaussian,hangleiter2022computational,grier2021complexity,quesada2018gaussian,deshpande2022quantum, martinez2024} and that finds interesting connections with properties of graphs~\cite{krenn2017quantum,gu2019quantum,quesada2019franck,yao2023design}. Unlike standard Boson Sampling~\cite{aaronson2011computational}, which requires hard-to-generate single-photon states~\cite{wang2019boson}, in GBS, a set of squeezed states are entangled by an interferometer and are then measured using photon-number~\cite{madsen2022quantum} or threshold detectors~\cite{zhong2021phase,zhong2020quantum,thekkadath2022experimental,yu2023universal,deng2023gaussian}. 
Note that the particle statistics of continous-variable quantum states also plays an important role in understanding issues in relativistic quantum information where harmonic oscillator detectors can be used to probe locally the properties of bosonic fields~\cite{caianiello1953quantum,friis2012motion,lapponi2023relativistic,papageorgiou2024particle,perche2023fully,torres2023particle}.

While it is by now well understood that the probabilities of Gaussian states measured with photon-number-resolving or threshold detectors are given by loop Hafnians~\cite{hamilton2017gaussian,kruse2019detailed,quesada2019franck,quesada2019simulating} and loop Torontonians~\cite{quesada2018gaussian,bulmer2022threshold} respectively, not so much is known for the moments of the photon-number distribution which have an infinite dimensional support. Unlike for distributions supported over bit strings (like the one associated with threshold detection), photon-number moments cannot be reduced to marginals~\cite{martnez2022classical}. Indeed, explicit results are only known for the first and second order moments and cumulants (also known as Ursell functions, connected correlation functions, truncated correlation functions or cluster functions~\cite{ursell_1927,duneau1973decrease}) of the photon-number distribution of Gaussian states~\cite{dodonov1994multidimensional,vallone2019means}.

In this contribution, we close this gap in the knowledge of the photon-number statistics of Gaussian states by deriving simple expressions as well as a graph representation for the moments and cumulants of this distribution. We also show that the calculation of both these expressions are $\#P$-hard and provide an exponential time algorithm for the calculation of cumulants. Besides fundamental interest in understanding the statistics of this family of states, our results can be used as verification tools for experimental implementations of GBS. Thus our work extends the theoretical proposal in Ref.~\cite{phillips2019benchmarking} where only second-order cumulant comparisons are  used as partial evidence in validating the correct functioning of GBS machines. Certifying the correct functioning of a GBS machine is still a significant open problem since like for many sampling-based quantum advantage proposals it is computationally intractable to calculate probabilities using the Born rule and experimentally infeasible to try to collect sufficient samples to have good frequentist estimates of these same quantities.

Using our graph representation, we are able to prove that, except for the photon-number mean, all odd-order cumulants are zero for multimode Gaussian states that have zero displacement and the property that the expectation value $\braket{\ah_i^\dagger \ah_j}$ is zero where the creation operator $\ah^\dagger_i$ and the destruction operator $\ah_j$ refer to two different modes.
This is the case if all the inputs of an interferometer with uniform losses are illuminated with the same single-mode zero-displacement Gaussian state. We obtain this result by introducing a new matrix function that we call the Montrealer. This matrix function could be of independent interest in the combinatorics/graph-theory community.

Moreover, we use the expressions we construct to compare the distribution of cumulants up to the fourth order for four different families of Gaussian states: squeezed, lossy squeezed, squashed and thermal states. While previous studies considered cumulants up to second order for squeezed and thermal states~\cite{phillips2019benchmarking}, our study goes further by looking at how hard it can be to distinguish squeezed and lossy squeezed against not only thermal states, but also squashed states. Unlike thermal states, squashed states possess phase-sensitive excess noise and are the optimal classical states (lacking any negativity of the Glauber-Sudarshan P-function) to spoof a lossy squeezed distribution~\cite{qi2020regimes,martnez2022classical}. 

This paper is organized as follows: we first review the Gaussian state formalism from which we develop a general expression for the photon-number moments in terms of the phase-insensitive and phase-sensitive moment matrices in Sec.~\ref{sec:review}. We then connect photon-number moments to the loop Hafnian in Sec.~\ref{sec:moments} and introduce their graph representation in Sec~\ref{sec:graph}. Having obtained expressions for the moments, we move on to the cumulants in Sec.~\ref{sec:cumulants}. In this section, we first motivate and derive from the concept of alternating walks new matrix functions, the Montrealer and the loop Montrealer, used to calculate photon-number cumulants of Gaussian states. Then, after exploring the properties of these new functions, we show their connection to Hamiltonian walks. We close this section by showing that the computation of moments and cumulants are $\#P$-hard problems and by providing an exponential time algorithm to calculate the cumulants. In Sec.~\ref{sec:numerics} we perform a statistical analysis comparing the cumulants up to fourth order for four different types of single-mode Gaussian states fed into a subset of the inputs of an interferometer. Conclusions are presented in Sec.~\ref{sec:fini}.
\section{Background on Gaussian states}\label{sec:review}
We consider a collection of $\ell$ modes that are described by annihilation $\ah$ and creation $\ch$ operators that we group as follows
\begin{align}
	\zh = \left(\ah_1,\dots, \ah\l, \ch_1, \dots, \ch\l\right)^T.
\end{align}
In terms of the components of $\zh$ we can write the bosonic canonical commutation relations as
\begin{align}
	[\hat{\zeta}_i, \hat{\zeta}_j^\dagger] = Z_{i,j} \text{ with } \bm{Z} = \begin{pmatrix}
		\I\l &  0_\ell \\
		0_\ell & - \I\l
	\end{pmatrix},
\end{align}
where $\I_\ell\,(0_\ell)$ is the identity (zero) matrix of size $\ell$.
It is customary to use the quadrature operators for the phase-space representation of modes. The necessary tools to go between the two representations are given in Appendix \ref{app:quadratures}.

The state of a quantum mechanical system is uniquely characterized by its density operator $\rho$. An equivalent complete description of a continuous-variable state is given in terms of its $s$-ordered characteristic function,
\begin{align}
	\chi_s(\z) = \Tr\left[\rho\hat{D}(\z)\right] \exp\left(\tfrac14 s ||\z||^2\right),
\end{align}
where $\z = (\a_1,\ldots, \a_\ell, \c_1,\ldots, \c_\ell)^T$ while the displacement operator is defined as
\begin{align}
	\hat{D}(\z) = \exp\left(\zh^\dagger \bm{Z} \z \right) = \exp\left( \sum_{i=1}^\ell \a_i \ch_i - \c_i \ah_i  \right).
\end{align}
Gaussian states are the set of states that have a Gaussian characteristic function in the variable $\z$ which we write as (cf. chapter 4 of Serafini ~\cite{serafini2017quantum} and Appendix A of  Ref.~\cite{thomas2021general}), 
\begin{align}\label{eq:chis}
	\chi_s(\z) = \exp{\left(-\tfrac{1}{2}  \z^\dagger \Z \S^{(s)}  \Z \z + \zb^\dagger \Z \z \right)},
\end{align}
where we introduce the vector of means and the hermitian covariance matrix as
\begin{align}
	\zb &= \braket{\zh}= \left(\ab_1,\dots,\ab\l,\cb_1,\dots,\cb\l\right)^T, \\
	\s_{ij}^{(s)} &= \tfrac{1}{2} \left[ \braket{\zeta_i \zeta_j\d +  \zeta_j\d \zeta_i} -s \ \delta_{ij}\right]-\bar{\zeta}_i\bar{\zeta}_j^{*} = \left[\s_{ji}^{(s)} \right]^*.
\end{align}
The vacuum state $\ket{\bm{0}}$, the unique state for which $a_i \ket{\bm{0}} = 0$, has $\bm{\Sigma}^{(s)} = \tfrac{1-s}{2} \mathbb{1}_{2\ell}$ and $\zb = \bm{0}$.

It is convenient to introduce the following block form of the covariance matrix
\begin{align}
	\S^{(s)} =  \tfrac{1-s}{2} \I_{2 \l} + \begin{pmatrix}
		\N^T  & \M \\
		\M^* & \N 
	\end{pmatrix},
\end{align}
where we use the (hermitian and positive-semidefinite) phase-insensitive matrix $\N$ and the (symmetric) phase-sensitive matrix $\M$ with entries
\begin{align}
	\n_{ij} &= \braket{\ch_i\ah_j}-\braket{\ch_i}\braket{\ah_j}=\n_{ji}^*,\\
	\m_{ij} &= \braket{\ah_i\ah_j}-\braket{\ah_i}\braket{\ah_j}=\m_{ji}.
\end{align}
The vacuum state has $\N = \M = 0_\ell$.

By setting $s=1$, we can obtain any normal-ordered moment of the creation and annihilation operators by taking derivatives \cite{gerry2005introductory,serafini2017quantum}
\begin{align}\label{eq:moments_anni_cre}
	&\braket{(\ch_1)^{r_1}\dots(\ch\l)^{r\l}\ah_1^{t_1}\dots\ah\l^{t\l}} \nonumber \\ &\quad =\left. \prod_{i=1}^{\ell}\left(\frac{\partial}{\partial\a_i}\right)^{r_i}\left(-\frac{\partial}{\partial\c_i}\right)^{t_i}\chi_{s=1}(\z) \right|_{\z = \bm{0}}.
\end{align}
By setting $s=-1$, one obtains anti-normal-ordered expectation values and similarly, for $s=0$, one obtains symmetric-ordered expectation values.

\section{Photon-number moments}\label{sec:moments}
To obtain any $s$-ordered moment of the creation and annihilation operators for a Gaussian state we need to evaluate
\begin{align}\label{eq:ders}
	\prod_{i=1}^{\ell}\left(\frac{\partial}{\partial\a_i}\right)^{r_i}&\left.\left(-\frac{\partial}{\partial\c_i}\right)^{t_i}\chi_s(\z)\right|_{\z=0}=\nonumber\\
	\prod_{i=1}^{\ell}\left(\frac{\partial}{\partial\a_i}\right)^{r_i}&\left(-\frac{\partial}{\partial\c_i}\right)^{t_i} \nonumber\\
	&\left.\exp{\left(\tfrac{1}{2}  \z^T \Z \A^{(s)} \Z \z + \zb^\dagger \Z \z \right)} \right|_{\z = \bm{0}},
\end{align}
where we define $\A^{(s)} = \X \S^{(s)}$, with block form
\begin{align}\label{eq:A}
	\A^{(s)} &=
	\begin{pmatrix}
		\M^* & \N+\tfrac{(1-s)}{2}\I\l \\ \N^T+\tfrac{(1-s)}{2}\I\l & \M
	\end{pmatrix}, \\
	\X &= \begin{pmatrix}
		\bm{0} & \I\l \\ \I\l & \bm{0}
	\end{pmatrix},
\end{align}
and we used $\z^{\dag}\X = \z^T$ and $\Z\X = -\X\Z$. We shall refer to $\A \equiv \A^{(s=1)}$ as the adjacency matrix of the Gaussian state. The previous result is shown in more detail in Ref. \cite{quesada2021beyond}. If in the last expression we swap $\c_i$ for $-\c_i$ (which we are allowed to do as they are dummy differentiation variables evaluated at zero), then $\left.\Z\z\right|_{\c_i\rightarrow -\c_i} = \z$ and we obtain
\begin{align}\label{eq:noders}
	&\prod_{i=1}^{\ell}\left.\left(\frac{\partial}{\partial\a_i}\right)^{r_i}\left(-\frac{\partial}{\partial\c_i}\right)^{t_i}\chi_s(\z)\right|_{\z = \bm{0}}=\\ &\quad 
	\prod_{i=1}^{\ell}\left(\frac{\partial}{\partial\a_i}\right)^{r_i}\left(\frac{\partial}{\partial\c_i}\right)^{t_i} \left.\exp{\left(\tfrac{1}{2}  \z^T \A^{(s)} \z + \zb^\dagger \z \right)} \right|_{\z = \bm{0}}. \nonumber 
\end{align}
Note that in going from Eq.~\eqref{eq:ders} to Eq.~\eqref{eq:noders} the minus signs in the derivatives with respect to $\alpha_i^*$ disappear together with the matrix $\bm{Z}$.
In Appendix A of Ref.~\cite{quesada2019simulating}, it is shown that the last expression coincides with the loop Hafnian allowing us to write
\begin{align}\label{eq:characteristic_lhaf}
	&\prod_{i=1}^{\ell}\left.\left(\frac{\partial}{\partial\a_i}\right)^{r_i}\left(-\frac{\partial}{\partial\c_i}\right)^{t_i}\chi_s(\z)\right|_{\z = \bm{0}} \nonumber \\
	&\quad =\lhaf\left[\fdiag\left(\A_{\bm{r}\oplus \bm{t}}^{(s)},\zb^*_{\bm{r}\oplus \bm{t}}\right)\right],
\end{align}
where $\bm{r}\oplus \bm{t} = \left(r_1,\dots,r\l,t_1,\dots,t\l\right)^T$. Note that in Appendix A of Ref.~\cite{quesada2019simulating} the argument of the loop hafnian corresponds to the precision matrix of the Husimi $Q$-function of a Gaussian state. Here instead, it is directly proportional to the $s-$ordered covariance matrix. 

In the equation above, the notation $\A_{\bm{k}}$ is used to define the matrix obtained by repeating the $i^{\text{th}}$ column and row of $\A$ a total of $k_i$ times. Similarly, the notation $\z_{\bm{k}}$ is used to represent the vector obtained by repeating the $i^{\text{th}}$ entry of $\z$ a total of $k_i$ times. $\fdiag(\A,\z)$ is the matrix obtained by replacing the diagonal elements of $\A$ by the entries of $\z$. Finally, $\lhaf\A$ is the loop Hafnian of matrix $\A$. Note that for zero-mean Gaussian states, that is if $\zb = \bm{0}$, one can replace the right-hand side of Eq.~\eqref{eq:characteristic_lhaf} by $\haf\A_{\bm{r}\oplus \bm{t}}^{(s)}$.

The Hafnian ($\haf$) of a symmetric matrix is a quantity that appears in graph theory. If we let $\bm{Q}$ be an arbitrary symmetric $(0,1)$-matrix representing the adjacency matrix of a given graph, then the Hafnian of $\bm{Q}$ gives the number of perfect matchings of that graph \cite{caianiello1953quantum}. The Hafnian of an arbitrary $m\times m$ symmetric matrix $\bm{Q}$ is given by \cite{bjorklund2019faster}
\begin{align}
	\haf \bm{Q} = \sum_{\bs\,\in\,\text{PMP}(m)}\prod_{(i,j)\,\in\,\bs}Q_{ij}\,,
\end{align}
where PMP($m$) stands for the set of perfect matching permutations of $m$ (even) objects~\cite{barvinok2016combinatorics}. Note that the expression above is equivalent to Isserlis'~\cite{isserlis1918formula} or Wick's~\cite{wick1950evaluation} theorem when the matrix in the argument of the hafnian is positive semi-definite and corresponds to the covariance matrix of a zero-mean multinormal distribution.

\begin{figure*}
	\captionsetup{justification = raggedright, singlelinecheck = false}
	\begin{center}
	\input{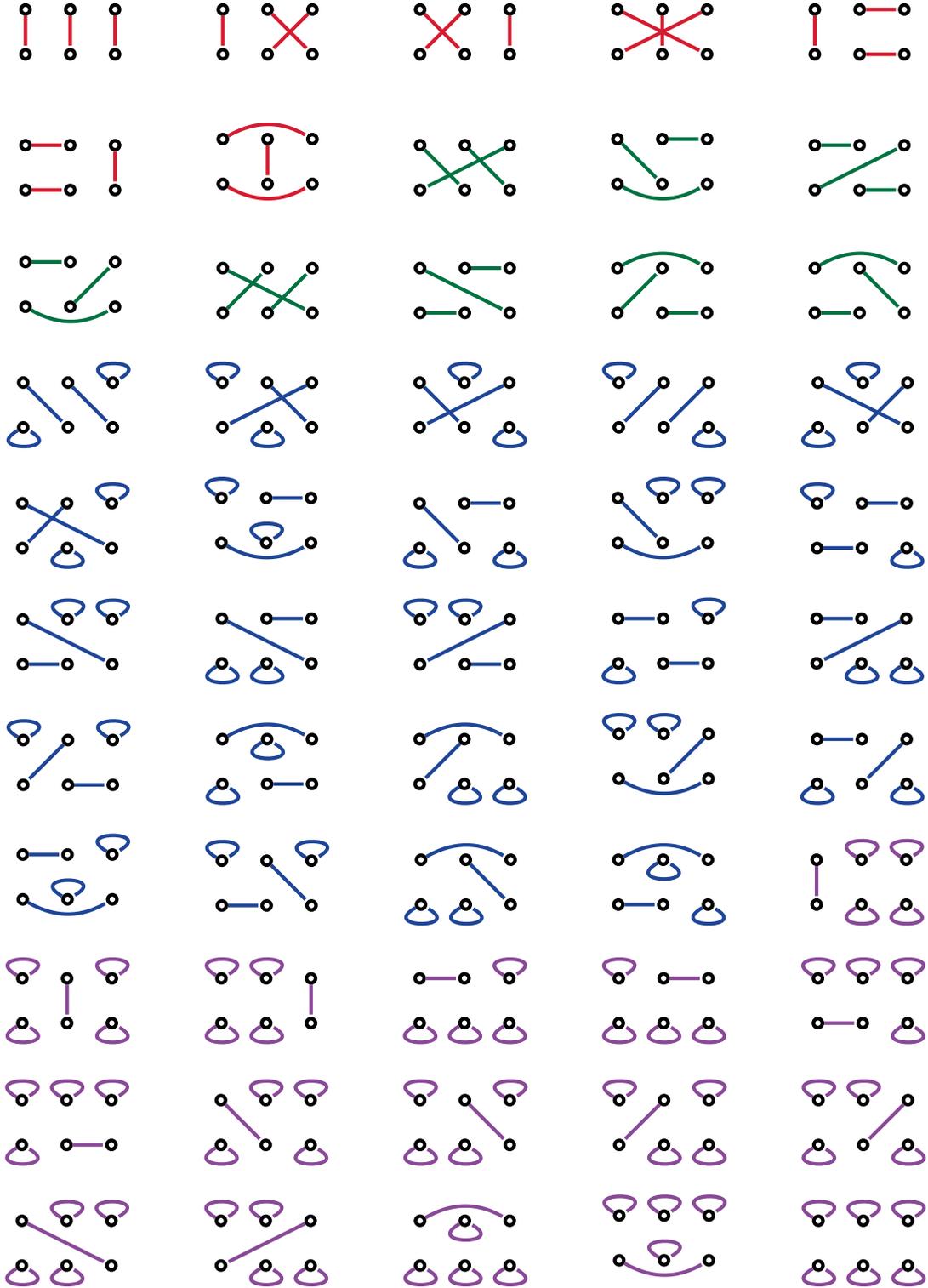}
	\end{center}
	\caption{\label{fig:lpm_3} Perfect matchings for a graph with six vertices allowing for loops. These perfect matchings correspond to the third order photon-number moment, $\braket{\nh_1\nh_2\nh_3}$.}
	
\end{figure*}

For $m=2,4$ one finds
\begin{align}
	\text{PMP}(2) &= \{(1,2)\}, \\
	\text{PMP}(4) &= \{(1,3)(2,4),(1,2)(3,4),(1,4)(2,3)\},
\end{align}
which correspond to the perfect matchings with edges coloured in red and green in Fig.~\ref{fig:lpm_1} and Fig.~\ref{fig:lpm_2} respectively. The set $\text{PMP}(6)$ is shown graphically in the graphs with vertices in red and green in Fig.~\ref{fig:lpm_3}.
Note that the number of elements in $\text{PMP}(2\ell)$ is given by $|\text{PMP}(2\ell)| = (2 \ell -1)!! = 1 \times 3 \times 5 \times \dots \times 2 \ell - 1.$

The loop Hafnian ($\lhaf$) is built similarly with one notable exception; it allows for repetition of indices. For an $m\times m$ symmetric matrix $\bm{Q}$, we have \cite{bjorklund2019faster}
\begin{align}
	\lhaf \bm{Q} = \sum_{\bs\,\in\,\text{SPM}(m)}\prod_{(i,j)\,\in\,\bs}Q_{ij}\,,
\end{align}
where, to account for the possibility of loops, the set of perfect matching permutations PMP is generalized to include single-pair matchings that we note SPM.
For $m=2,4$ one finds
\begin{align}
	\text{SPM}(2) =& \{(1,2),(1)(2)\}, \\
	\text{SPM}(4) =& \{(1,3)(2,4),(1,2)(3,4),(1,4)(2,3),\\
	&(1,2)(3)(4),(1)(2)(3,4),(1,4)(2)(3), \nonumber \\
	&(1)(2,3)(4),(1)(2,4)(3),(1,2)(3)(4), \nonumber \\
	&(1)(2)(3)(4)\}, \nonumber
\end{align}
which correspond to the perfect matchings in Fig.~\ref{fig:lpm_1} and Fig.~\ref{fig:lpm_2} respectively.  The set $\text{SPM}(6)$ is shown graphically in Fig.~\ref{fig:lpm_3}. The number of elements in $\text{SPM}(2\ell)$ is given by $|\text{SPM}(2\ell)| = T(2\ell)$ where $T(n)$ is the $n^{\text{th}}$ Telephone number~\cite{bjorklund2019faster}. Note that for large number of vertices there are superpolynomially many more single-pair matchings than perfect matchings since $T(n) / (n-1)!! \sim \tfrac12 \exp\left(\sqrt{n} - \tfrac14 \right)$ as $n \to \infty$ ~\cite{bjorklund2019faster}.  Also the loop Hafnian and the Hafnian of matrix $\bm{Q}$ coincides in the case where $\bm{Q}$ has zeros along its diagonal.

While the characteristic function of a Gaussian state has been reported in the literature \cite{phillips2019benchmarking}, to the best of our knowledge, no explicit connection has been made between moments of the creation and annihilation operators and the loop Hafnian. This connection is useful since it allows us to use recently developed fast algorithms to calculate loop Hafnians. Indeed, using the results from Bulmer et al.~\cite{bulmer2022boundary}, the loop Hafnian in Eq.~\eqref{eq:characteristic_lhaf} can be calculated in time that scales as $\mathcal{O}\left(c^3\sqrt{\prod_{i=1}^{\ell}(r_i+1)(t_i+1)}\right)$ where $c$ is the number of non-zero entries in $\bm{r}\oplus\bm{t}$. In particular if for all $\ell$ modes, $r_i = t_i=1$, then the time complexity will scale as $\ell^3 2^\ell$.

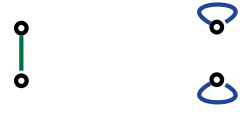
\begin{figure}[!t]
	\captionsetup{justification = raggedright, singlelinecheck = false}
	\begin{centering}\begin{tikzpicture}[scale=.7, shorten >=1pt, auto, node distance=1cm, ultra thick]
    \tikzstyle{node_style} = [circle,draw=black, inner sep=0pt, minimum size=4pt]
    \tikzstyle{edge_style} = [-,draw=white, line width=2, thick, dashed]
    \tikzstyle{edge_styleg} = [-,draw=mtl_green , line width=2, ultra thick]

    \node[node_style] (v1) at (0,1) {};
    \node[node_style] (v2) at (0,0) {};

    \draw[edge_styleg]  (v1) edge (v2);

    \draw[edge_style]  (v1) to [loop right, in=30,out=150,looseness=10] (v1);
    \draw[edge_style]  (v2) to [loop right, in=-30,out=-150,looseness=10] (v2);

    \end{tikzpicture} \vspace{0 cm}\hspace{1 cm}
\
\begin{tikzpicture}[scale=.7, shorten >=1pt, auto, node distance=1cm, ultra thick]
    \tikzstyle{node_style} = [circle,draw=black, inner sep=0pt, minimum size=4pt]
    \tikzstyle{edge_style} = [-,draw=white, line width=2, thick, dashed]
    \tikzstyle{edge_styleg} = [-,draw=mtl_blue , line width=2, ultra thick]

    \node[node_style] (v1) at (0,1) {};
    \node[node_style] (v2) at (0,0) {};

    \draw[edge_styleg]  (v1) to [loop right, in=30,out=150,looseness=10] (v1);
    \draw[edge_styleg]  (v2) to [loop right, in=-30,out=-150,looseness=10] (v2);

    \end{tikzpicture} \vspace{0 cm}\hspace{1 cm}
    \end{centering}
	\caption{\label{fig:lpm_1} Perfect matchings for a graph with two vertices allowing for loops. These perfect matchings correspond to the first order photon-number moment, $\braket{\nh_1}$.}
\end{figure}
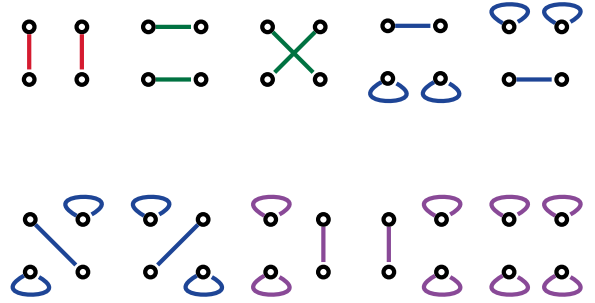
\begin{figure}[!t]
	\captionsetup{justification = raggedright, singlelinecheck = false}
	\begin{tikzpicture}[scale=.7, shorten >=1pt, auto, node distance=1cm, ultra thick]
    \tikzstyle{node_style} = [circle,draw=black, inner sep=0pt, minimum size=4pt]
    \tikzstyle{edge_style} = [-,draw=white, line width=2, thick, dashed]
    \tikzstyle{edge_styleg} = [-,draw=mtl_red , line width=2, ultra thick]

    \node[node_style] (v1) at (0,1) {};
    \node[node_style] (v2) at (1,1) {};
    \node[node_style] (v3) at (0,0) {};
    \node[node_style] (v4) at (1,0) {};

    \draw[edge_styleg]  (v1) edge (v3);
    \draw[edge_styleg]  (v2) edge (v4);

    \draw[edge_style]  (v2) to [loop right, in=30,out=150,looseness=10] (v2);
    \draw[edge_style]  (v3) to [loop right, in=-30,out=-150,looseness=10] (v3);

    \end{tikzpicture} \vspace{0.12 cm}\hspace{-0.65 cm}
\
\begin{tikzpicture}[scale=.7, shorten >=1pt, auto, node distance=1cm, ultra thick]
    \tikzstyle{node_style} = [circle,draw=black, inner sep=0pt, minimum size=4pt]
    \tikzstyle{edge_style} = [-,draw=white, line width=2, thick, dashed]
    \tikzstyle{edge_styleg} = [-,draw=mtl_green , line width=2, ultra thick]

    \node[node_style] (v1) at (0,1) {};
    \node[node_style] (v2) at (1,1) {};
    \node[node_style] (v3) at (0,0) {};
    \node[node_style] (v4) at (1,0) {};

    \draw[edge_styleg]  (v1) edge (v2);
    \draw[edge_styleg]  (v3) edge (v4);

    \draw[edge_style]  (v2) to [loop right, in=30,out=150,looseness=10] (v2);
    \draw[edge_style]  (v3) to [loop right, in=-30,out=-150,looseness=10] (v3);

    \end{tikzpicture} \vspace{0.12 cm}\hspace{-0.65 cm}
\
\begin{tikzpicture}[scale=.7, shorten >=1pt, auto, node distance=1cm, ultra thick]
    \tikzstyle{node_style} = [circle,draw=black, inner sep=0pt, minimum size=4pt]
    \tikzstyle{edge_style} = [-,draw=white, line width=2, thick, dashed]
    \tikzstyle{edge_styleg} = [-,draw=mtl_green , line width=2, ultra thick]

    \node[node_style] (v1) at (0,1) {};
    \node[node_style] (v2) at (1,1) {};
    \node[node_style] (v3) at (0,0) {};
    \node[node_style] (v4) at (1,0) {};

    \draw[edge_styleg]  (v1) edge (v4);
    \draw[edge_styleg]  (v2) edge (v3);

    \draw[edge_style]  (v2) to [loop right, in=30,out=150,looseness=10] (v2);
    \draw[edge_style]  (v3) to [loop right, in=-30,out=-150,looseness=10] (v3);

    \end{tikzpicture} \vspace{0.12 cm}\hspace{-0.65 cm}
\
\begin{tikzpicture}[scale=.7, shorten >=1pt, auto, node distance=1cm, ultra thick]
    \tikzstyle{node_style} = [circle,draw=black, inner sep=0pt, minimum size=4pt]
    \tikzstyle{edge_style} = [-,draw=white, line width=2, thick, dashed]
    \tikzstyle{edge_styleg} = [-,draw=mtl_blue , line width=2, ultra thick]

    \node[node_style] (v1) at (0,1) {};
    \node[node_style] (v2) at (1,1) {};
    \node[node_style] (v3) at (0,0) {};
    \node[node_style] (v4) at (1,0) {};

    \draw[edge_styleg]  (v1) edge (v2);
    \draw[edge_styleg]  (v3) to [loop right, in=-30,out=-150,looseness=10] (v3);
    \draw[edge_styleg]  (v4) to [loop right, in=-30,out=-150,looseness=10] (v4);
    
    \draw[edge_style]  (v2) to [loop right, in=30,out=150,looseness=10] (v2);

    \end{tikzpicture} \vspace{0.12 cm}\hspace{-0.65 cm}
\
\begin{tikzpicture}[scale=.7, shorten >=1pt, auto, node distance=1cm, ultra thick]
    \tikzstyle{node_style} = [circle,draw=black, inner sep=0pt, minimum size=4pt]
    \tikzstyle{edge_style} = [-,draw=white, line width=2, thick, dashed]
    \tikzstyle{edge_styleg} = [-,draw=mtl_blue , line width=2, ultra thick]

    \node[node_style] (v1) at (0,1) {};
    \node[node_style] (v2) at (1,1) {};
    \node[node_style] (v3) at (0,0) {};
    \node[node_style] (v4) at (1,0) {};

    \draw[edge_styleg]  (v3) edge (v4);
    \draw[edge_styleg]  (v1) to [loop right, in=30,out=150,looseness=10] (v1);
    \draw[edge_styleg]  (v2) to [loop right, in=30,out=150,looseness=10] (v2);
    
    \draw[edge_style]  (v4) to [loop right, in=-30,out=-150,looseness=10] (v4);

    \end{tikzpicture} \vspace{0.12 cm}\hspace{-0.65 cm}

\begin{tikzpicture}[scale=.7, shorten >=1pt, auto, node distance=1cm, ultra thick]
    \tikzstyle{node_style} = [circle,draw=black, inner sep=0pt, minimum size=4pt]
    \tikzstyle{edge_style} = [-,draw=white, line width=2, thick, dashed]
    \tikzstyle{edge_styleg} = [-,draw=mtl_blue , line width=2, ultra thick]

    \node[node_style] (v1) at (0,1) {};
    \node[node_style] (v2) at (1,1) {};
    \node[node_style] (v3) at (0,0) {};
    \node[node_style] (v4) at (1,0) {};

    \draw[edge_styleg]  (v1) edge (v4);
    \draw[edge_styleg]  (v2) to [loop right, in=30,out=150,looseness=10] (v2);
    \draw[edge_styleg]  (v3) to [loop right, in=-30,out=-150,looseness=10] (v3);

    \end{tikzpicture} \vspace{0.12 cm}\hspace{-0.65 cm}
\
\begin{tikzpicture}[scale=.7, shorten >=1pt, auto, node distance=1cm, ultra thick]
    \tikzstyle{node_style} = [circle,draw=black, inner sep=0pt, minimum size=4pt]
    \tikzstyle{edge_style} = [-,draw=white, line width=2, thick, dashed]
    \tikzstyle{edge_styleg} = [-,draw=mtl_blue , line width=2, ultra thick]

    \node[node_style] (v1) at (0,1) {};
    \node[node_style] (v2) at (1,1) {};
    \node[node_style] (v3) at (0,0) {};
    \node[node_style] (v4) at (1,0) {};

    \draw[edge_styleg]  (v2) edge (v3);
    \draw[edge_styleg]  (v1) to [loop right, in=30,out=150,looseness=10] (v1);
    \draw[edge_styleg]  (v4) to [loop right, in=-30,out=-150,looseness=10] (v4);

    \end{tikzpicture} \vspace{0.12 cm}\hspace{-0.65 cm}
\
\begin{tikzpicture}[scale=.7, shorten >=1pt, auto, node distance=1cm, ultra thick]
    \tikzstyle{node_style} = [circle,draw=black, inner sep=0pt, minimum size=4pt]
    \tikzstyle{edge_style} = [-,draw=white, line width=2, thick, dashed]
    \tikzstyle{edge_styleg} = [-,draw=mtl_purple , line width=2, ultra thick]

    \node[node_style] (v1) at (0,1) {};
    \node[node_style] (v2) at (1,1) {};
    \node[node_style] (v3) at (0,0) {};
    \node[node_style] (v4) at (1,0) {};

    \draw[edge_styleg]  (v2) edge (v4);
    \draw[edge_styleg]  (v1) to [loop right, in=30,out=150,looseness=10] (v1);
    \draw[edge_styleg]  (v3) to [loop right, in=-30,out=-150,looseness=10] (v3);

    \draw[edge_style]  (v4) to [loop right, in=-30,out=-150,looseness=10] (v4);

    \end{tikzpicture} \vspace{0.12 cm}\hspace{-0.65 cm}
\
\begin{tikzpicture}[scale=.7, shorten >=1pt, auto, node distance=1cm, ultra thick]
    \tikzstyle{node_style} = [circle,draw=black, inner sep=0pt, minimum size=4pt]
    \tikzstyle{edge_style} = [-,draw=white, line width=2, thick, dashed]
    \tikzstyle{edge_styleg} = [-,draw=mtl_purple , line width=2, ultra thick]

    \node[node_style] (v1) at (0,1) {};
    \node[node_style] (v2) at (1,1) {};
    \node[node_style] (v3) at (0,0) {};
    \node[node_style] (v4) at (1,0) {};

    \draw[edge_styleg]  (v1) edge (v3);
    \draw[edge_styleg]  (v2) to [loop right, in=30,out=150,looseness=10] (v2);
    \draw[edge_styleg]  (v4) to [loop right, in=-30,out=-150,looseness=10] (v4);

    \draw[edge_style]  (v1) to [loop right, in=30,out=150,looseness=10] (v1);
    
    \end{tikzpicture} \vspace{0.12 cm}\hspace{-0.65 cm}
\
\begin{tikzpicture}[scale=.7, shorten >=1pt, auto, node distance=1cm, ultra thick]
    \tikzstyle{node_style} = [circle,draw=black, inner sep=0pt, minimum size=4pt]
    \tikzstyle{edge_style} = [-,draw=white, line width=2, thick, dashed]
    \tikzstyle{edge_styleg} = [-,draw=mtl_purple , line width=2, ultra thick]

    \node[node_style] (v1) at (0,1) {};
    \node[node_style] (v2) at (1,1) {};
    \node[node_style] (v3) at (0,0) {};
    \node[node_style] (v4) at (1,0) {};

    \draw[edge_styleg]  (v1) to [loop right, in=30,out=150,looseness=10] (v1);
    \draw[edge_styleg]  (v2) to [loop right, in=30,out=150,looseness=10] (v2);
    \draw[edge_styleg]  (v3) to [loop right, in=-30,out=-150,looseness=10] (v3);
    \draw[edge_styleg]  (v4) to [loop right, in=-30,out=-150,looseness=10] (v4);

    \end{tikzpicture} \vspace{0.12 cm}\hspace{-0.65 cm}
    
	\caption{\label{fig:lpm_2} Perfect matchings for a graph with four vertices allowing for loops. These perfect matchings correspond to the second order photon-number moment, $\braket{\nh_1 \nh_2}$.}
\end{figure}
Given the previous results, we can develop an explicit equation linking the photon-number moments to the loop Hafnian. The explicit expression giving normal-ordered expansion of the powers of the photon-number operator~\cite{blasiak2007combinatorics} is 
\begin{align}\label{eq:ordered}
	(\ch\ah)^m &= \sum_{n=1}^{m} \left\{\begin{matrix} m \\n  \end{matrix} \right\} 
	\left(\ch\right)^n\ah^n, \text{ where,} \\
	\left\{\begin{matrix} m \\n  \end{matrix} \right\} &= \sum_{k=1}^n{n\choose k}\frac{k^m}{n!}(-1)^{n-k},
\end{align}
is a Stirling number of the second kind.

With this expression, we can write the first main result of our work giving expectation values of products of  power of number operator as follows (recall that $\A\equiv \A^{(s=1)}$),
\begin{align}\label{eq:moments_general}
	&\braket{\nh_1^{r_1}\dots\nh\l^{r\l}} = \\
	&\sum_{j_1=1}^{r_1}\dots\sum_{j\l=1}^{r\l}
	\left\{\begin{matrix} r_1 \\j_1  \end{matrix} \right\} \dots
	\left\{\begin{matrix} r\l \\j\l  \end{matrix} \right\}
	\lhaf\left[\fdiag\left(\A_{\js\oplus \js},\zb^*_{\js\oplus \js}\right)\right].  \nonumber
\end{align}
For zero-mean Gaussian states, the formula above reduces to
\begin{align}\label{eq:moments_no_displacement}
	\braket{\nh_1^{r_1}\dots\nh\l^{r\l}} = \sum_{j_1=1}^{r_1}\dots\sum_{j\l=1}^{r\l} \left\{\begin{matrix} r_1 \\j_1  \end{matrix} \right\} \dots
	\left\{\begin{matrix} r\l \\j\l  \end{matrix} \right\}\haf\A_{\js\oplus \js}.
\end{align}
Note that in general the problem of computing the Hafnian (and loop Hafnian) even for generic (0,1)-matrices has been shown to be $\#P-$complete~\cite{barvinok2016combinatorics}, however the matrices appearing inside the moment expression are not arbitrary since they are linked to covariance matrices satisfying the uncertainty relation (cf. Appendix~\ref{app:quadratures}). However, as we show below,  at least for a subset of the Gaussian states the calculation of their moments is $\#P-$ hard. To this end, note that in the case where $\z = \M = \bm{0}$, $r_i=1$ for all $i$  and thus $\A = \left( \begin{smallmatrix} 0_\ell  & \N \\ \N^T & 0_\ell
\end{smallmatrix} \right)$ we obtain the photon-number moments of thermal states in terms of Permanents
\begin{align}\label{eq:moments_Permanent}
	\left. \braket{\nh_1\dots\nh\l} \right|_{\M = \z = 0} =\per \N
\end{align}
where $\per \A = \haf \left( \begin{smallmatrix} 0_\ell  & \A \\ \A^T & 0_\ell
\end{smallmatrix} \right) = \sum_{\sigma \in S_\ell} \prod_{ i=1}^\ell A_{i,\sigma(i)} $ for $\A$ a square matrix of dimension $\ell$ and $S_\ell$ the symmetric group.
This observation allows us to state that the calculation of photon-number moments of Gaussian states, at least when there are no repeated photon-number operators and thus $\bm{r}=(1,\ldots,1)$, is $\#P$-hard since, as shown by Grier and Schaeffer~\cite{grier2016new}, Permanents of positive-semidefinite matrices are $\#P$-hard to compute.

\begin{figure}
	\captionsetup{justification = raggedright, singlelinecheck = false}
	\includegraphics[width=0.48\textwidth]{"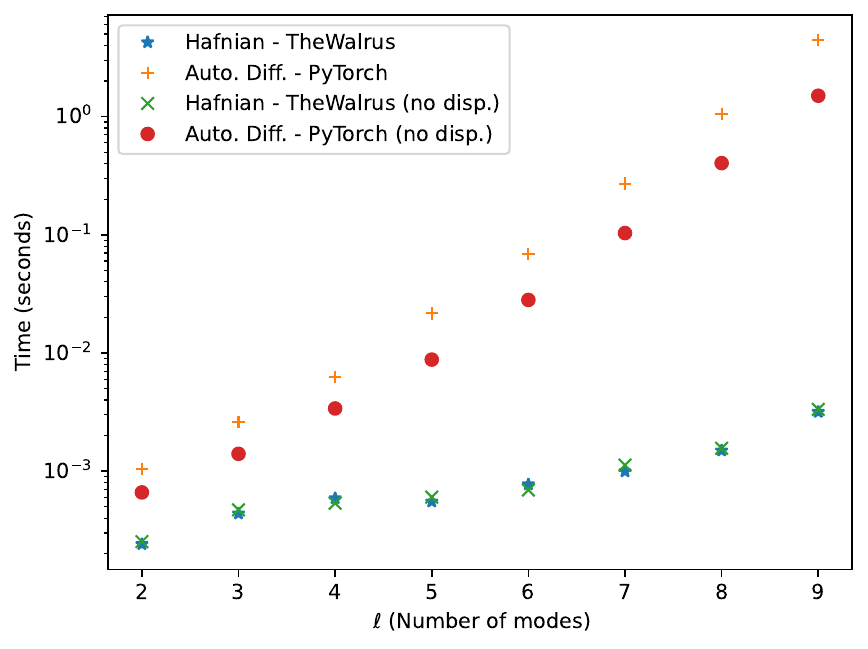"}
	\caption{\label{fig:timing_moments}We benchmarked the time required to calculate photon-number moment of $\ell$ modes by either using the Hafnian algorithms of \texttt{thewalrus}~\cite{gupt2019walrus} or using automatic differentiation on the generating functions using \texttt{pytorch}~\cite{torch} using the methods from Fitzke et al.~\cite{fitzke2022generating}. The  code used to perform the simulation together with hardware and software details is publicly available at~\cite{polyq2023bench}. We display results for both states with and without displacements. Note that this has a significant effect on the automatic differentiation timings, but leaves largely unchanged the corresponding timings for Hafnian methods.}
\end{figure}

We conclude this section by writing the moment generating function for the photon-number statistics of Gaussian states. This is defined as
\begin{align}\label{eq:trace}
	\mathcal{M}(\bm{t}) = \braket{e^{\hat{\bm{n}}\cdot \bm{t}}} =& \text{Tr}\left[ \rho(\bm{\S},\z) e^{\hat{\bm{n}}\cdot \bm{t}} \right]\nonumber \\   =& Z(-\bm{t}) \text{Tr} \left[ \rho(\bm{\S},\z) \tau(-\bm{t})  \right],
\end{align}
where we have written $\rho(\bm{\S},\z)$ to emphasize the specific Gaussian state in consideration, the vector $\bm{t}\in \mathbb{R}^\ell$ and $\hat{\bm{n}}^T = (\nh_1,\ldots,\nh_\ell)$  and we introduced the normalized Gibbs state
\begin{align}
	\tau(\bm{\beta}) = \frac{e^{-\hat{\bm{n}}\cdot \bm{\beta}} }{Z(\bm{\beta})}, \quad Z(\bm{\beta}) = \text{Tr}\left( e^{-\hat{\bm{n}}\cdot \bm{\beta}} \right) = \prod_{i=1}^\ell  \frac{1}{1-e^{-\beta_i}}.
\end{align}
The trace in Eq.~\eqref{eq:trace} can now be evaluated using phase-space methods by writing the Gaussian $Q$ -function of the state $\rho(\bm{\S},\z)$ and the Gaussian $P$-function of the state $\tau(\bm{\beta})$ to obtain 
\begin{align}\label{eq:M}
	\mathcal{M}(\bm{t}) =  \frac{\exp\left( \frac{1}{2} \bar{\bm{\zeta}}^\dagger \bm{G} [\I_{2 \ell} - \bm{G} \bm{\Sigma}]^{-1} \bar{\bm{\zeta}}\right)}{\sqrt{\det\left( \I_{2 \ell} - \bm{G} \bm{\Sigma}  \right)}}\text{, where} \\
	\bm{G} = \left[ \oplus_{i=1}^\ell (e^{t_i}-1) \right] \oplus \left[ \oplus_{i=1}^\ell (e^{t_i}-1) \right].
\end{align}
As a sanity check, it is easy to confirm that in the case where $\S = 0_{2 \ell}$, which corresponds to a product of coherent states in each mode, the moment-generating function reduces to a product of moment generating functions for Poisson distributions with means $\lambda_i = |\alpha_i|^2$
\begin{align}
	\mathcal{M}(\bm{t}) |_{\bm{\Sigma} \to  0 } = \prod_{i=1}^\ell\exp\left(  \lambda_i (e^{t_i}-1) \right).
\end{align}
With Eq.~\eqref{eq:M} at hand one can obtain moments by taking derivatives with respect to $\bm{t}$ 
\begin{align}\label{eq:MM}
	\braket{\nh_1^{r_1}\dots\nh\l^{r\l}} = \left. \prod_{i=1}^{\ell} \left(\frac{\partial}{\partial t_i}\right)^{r_i} \mathcal{M}(\bm{t}) \right|_{\bm{t} = \bm{0}}.
\end{align}
As suggested in Ref.~\cite{fitzke2022generating} one can use automatic differentiation techniques to obtain moments.  We compare their methods with ours and find that it is significantly faster to use the algorithms in Refs.~\cite{bulmer2022boundary,bjorklund2019faster} to calculate Hafnians to obtain moments. Indeed, as seen in Fig.~\ref{fig:timing_moments}, the Hafnian methods are significantly faster, even for small numbers of modes, and moreover have a better scaling as the problem size is increased.

\section{Graph representation of photon-number moments}\label{sec:graph}
\begin{figure}[!t]
	\captionsetup{justification = raggedright, singlelinecheck = false}
	\begin{tikzpicture}[scale=.7, shorten >=1pt, auto, node distance=1cm, ultra thick]
    \tikzstyle{node_style} = [circle,draw=black, inner sep=0pt, minimum size=14pt]
    \tikzstyle{text_style} = [circle, inner sep=0pt, minimum size=14pt, font = \Large]
    \tikzstyle{edge_style} = [-,draw=white, line width=2, thick, dashed]
    \tikzstyle{edge_styleg} = [-,draw=mtl_blue , line width=2, ultra thick]

    \node[node_style] (v1) at (0,2) {1};
    \node[node_style] (v2) at (2,2) {2};
    \node[node_style] (v3) at (4,2) {3};
    \node[node_style] (v4) at (6,2) {4};
    \node[node_style] (v5) at (8,2) {5};
    \node[node_style] (v6) at (0,0) {6};
    \node[node_style] (v7) at (2,0) {7};
    \node[node_style] (v8) at (4,0) {8};
    \node[node_style] (v9) at (6,0) {9};
    \node[node_style] (v10) at (8,0) {10};

    \draw[edge_styleg]  (v1) edge (v7);
    \draw[edge_styleg]  (v2) edge (v3);
    \draw[edge_styleg]  (v4) edge (v8);
    \draw[edge_styleg]  (v6) edge [bend right] node[below] {} (v9);
    \draw[edge_styleg]  (v5) to [loop right, in=30,out=150,looseness=10] (v5);
    \draw[edge_styleg]  (v10) to [loop right, in=-30,out=-150,looseness=10] (v10);

    \node[text_style] at (3,2.7) {$A_{2,3}$};
    \node[text_style] at (0.7,-1.2) {$A_{6,9}$};
    \node[text_style] at (1.9,1) {$A_{1,7}$};
    \node[text_style] at (5.9,1) {$A_{4,8}$};
    \node[text_style] at (6.5,3.2) {$\bar{\zeta}_{5}^{*}$};
    \node[text_style] at (6.5,-1.2) {$\bar{\zeta}_{10}^{*}$};
    
    \end{tikzpicture}\hspace{-2 cm}

\begin{tikzpicture}[scale=.7, shorten >=1pt, auto, node distance=1cm, ultra thick]

    \tikzstyle{node_style} = [circle,draw=white, inner sep=0pt, minimum size=14pt]
    \tikzstyle{text_style} = [circle,draw=white, inner sep=0pt, minimum size=14pt]
    \tikzstyle{edge_arrow} = [->,draw=black , line width=12, ultra thick]

    \node[node_style] (v1) at (0,2) {};
    \node[node_style] (v2) at (2,2) {};
    \node[node_style] (v3) at (4,2) {};
    \node[node_style] (v4) at (6,2) {};
    \node[node_style] (v5) at (8,2) {};
    \node[node_style] (v6) at (0,0) {};
    \node[node_style] (v7) at (2,0) {};
    \node[node_style] (v8) at (4,0) {};
    \node[node_style] (v9) at (6,0) {};
    \node[node_style] (v10) at (8,0) {};
    
    \draw[edge_arrow]  (v3) edge (v8);

    \end{tikzpicture}\hspace{-0.75 cm}

\captionsetup{justification = raggedright, singlelinecheck = false}
\begin{tikzpicture}[scale=.7, shorten >=1pt, auto, node distance=1cm, ultra thick]
    \tikzstyle{node_style} = [circle,draw=black, inner sep=0pt, minimum size=14pt]
    \tikzstyle{text_style} = [circle, inner sep=0pt, minimum size=14pt, font = \Large]
    \tikzstyle{edge_style} = [-,draw=white, line width=2, thick, dashed]
    \tikzstyle{edge_styleg} = [-,draw=mtl_blue , line width=2, ultra thick]

    \node[node_style] (v1) at (0,2) {1};
    \node[node_style] (v2) at (2,2) {2};
    \node[node_style] (v3) at (4,2) {3};
    \node[node_style] (v4) at (6,2) {4};
    \node[node_style] (v5) at (8,2) {5};
    \node[node_style] (v6) at (0,0) {6};
    \node[node_style] (v7) at (2,0) {7};
    \node[node_style] (v8) at (4,0) {8};
    \node[node_style] (v9) at (6,0) {9};
    \node[node_style] (v10) at (8,0) {10};

    \draw[edge_styleg]  (v1) edge (v7);
    \draw[edge_styleg]  (v2) edge (v3);
    \draw[edge_styleg]  (v4) edge (v8);
    \draw[edge_styleg]  (v6) edge [bend right] node[below] {} (v9);
    \draw[edge_styleg]  (v5) to [loop right, in=30,out=150,looseness=10] (v5);
    \draw[edge_styleg]  (v10) to [loop right, in=-30,out=-150,looseness=10] (v10);

    \node[text_style] at (3,2.7) {$M_{2,3}^{*}$};
    \node[text_style] at (0.7,-1.2) {$M_{1,4}$};
    \node[text_style] at (1.9,1) {$N_{1,2}$};
    \node[text_style] at (6.1,1) {$N_{3,4}^{*}$};
    \node[text_style] at (6.5,3.2) {$\bar{\alpha}_{5}^{*}$};
    \node[text_style] at (6.5,-1.2) {$\bar{\alpha}_{5}$};
    
    \end{tikzpicture}\hspace{-2 cm}
	\caption{\label{fig:graph_legend} Graph representation of $\lhaf\left[\fdiag\left(\A,\zb^*\right)\right]$ perfect matchings for $\ell=5$. The graph is made of two rows of $\ell$ vertices. Every vertex represents a row or a column from $\fdiag\left(\A,\zb^*\right)$ while the edges represent its elements. The perfect matchings of the loop Hafnian associated with a graph is obtained from the product of the matrix elements associated with the edges.}
\end{figure}
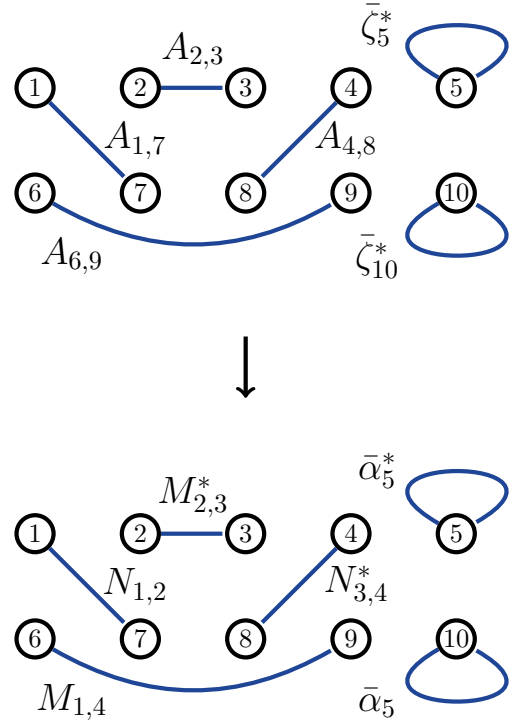

Up to this point we have not explicitly used the block structure of the adjacency matrices needed to obtain moments. Note that these matrices are obtained from quantum covariance matrices that need to satisfy the uncertainty relation (cf. Appendix~\ref{app:quadratures}). Indeed, these matrices cannot be completely arbitrary since their Hafnians need to be non-negative as they correspond to moments of a distribution with support over the non-negative integers.
In order to facilitate the visualisation of the photon-number moments of Eq.~\eqref{eq:moments_general}, we introduce the graph representation illustrated on Fig.~\ref{fig:graph_legend}. Without loss of generality, we will from now on work with $\A$ and $\z$ instead of $\A_{\js\oplus \js}$ and $\z_{\js\oplus \js}$. Equivalently, we assume that $r_i=1\,\forall\,i\,\in\,\{1,\dots,\ell\}$. The general case for which $\bm{r}$ allows for null values can be reduced to this simpler case by removing from the adjacency matrix the rows and columns for which $r_i=0$. The cases for which $r_i>1$ can still be handled by taking higher-order derivatives in Eq.~\eqref{eq:MM} but are no longer homogeneous polynomials in the entries of $\M$ and $\N$ making a graph theoretic interpretation more complicated.

The graph representing the Gaussian state is made of two rows of $\ell$ vertices numbered from $1$ to $2\ell$. The top row contains the vertices $\{1,2,\ldots,\ell\}$, while the bottom row contains the vertices $\{\ell+1,\ell+2,\ldots,2\ell\}$. Every vertex represents a matrix index while edges represent elements of $\fdiag\left(\A,\zb^*\right)$. For example, an edge between vertices $1$ and $7$ refers to matrix element $A_{17}$, while a loop on vertex $5$ refers to the fifth element of the conjugated vector of means $\zb^*$. Given the block form of the adjacency matrix in Eq.~\eqref{eq:A}, every edge on this graph can be directly expressed in terms of phase-sensitive and phase-insensitive matrix elements. An edge connecting two vertices in the bottom row is associated with an element of the phase-sensitive matrix $\M$ while an edge connecting vertices on the top row is associated with $\M^*$. The edges connecting vertices in opposite rows crossing from left to right are associated with the phase-insensitive matrix $\N$ while the edges crossing right to left are linked to elements of $\N^*$. Finally, note that loops in the bottom (top) row correspond to (conjugated) displacements $\bar{\bm{\alpha}}\,(\bar{\bm{\alpha}}^*)$. 

\begin{table}[]
	\begin{tabular}{lllll}
		\hline
		loop Hafnian & \crule[mtl_green]{1cm}{0.3cm} & \crule[mtl_red]{1cm}{0.3cm} & \crule[mtl_blue]{1cm}{0.3cm} & \crule[mtl_purple]{1cm}{0.3cm}\\
		Hafnian & \crule[mtl_green]{1cm}{0.3cm} & \crule[mtl_red]{1cm}{0.3cm} & &  \\
		loop Montrealer & \crule[mtl_green]{1cm}{0.3cm} & \crule[mtl_blue]{1cm}{0.3cm} & & \\
				Montrealer & \crule[mtl_green]{1cm}{0.3cm} & & & \\
		\hline
	\end{tabular}
	\captionsetup{justification = raggedright, singlelinecheck = false}
	\caption{Color coding used for the graph representation of  loop Hafnians, Hafnians, loop Montrealers and Montrealers.}
	\label{tab:color_legend}
\end{table}

\section{Photon-number cumulants}\label{sec:cumulants}
A probability distribution is entirely specified by its cumulants which are the genuine correlations between subsets of random variables~\cite{ursell_1927,fisher1932derivation}. The first three cumulants correspond with the first three central moments, but this is not true for fourth order and beyond. We write the photon-number cumulants as
\begin{align} \label{eq:cumulant_general}
	&\langle\langle\nh_{\gamma_1}\dots \nh_{\gamma_m}\rangle\rangle = \nonumber \\
	&\quad\sum_{\pi\,\in\, P(\gs)}\left(|\pi|-1\right)!(-1)^{|\pi|-1}\prod_{\bs\,\in\,\pi}\left\langle\prod_{i\,\in\,\bs}\nh_i\right\rangle,
\end{align}
where $\gs$ is a collection of mode labels. We use $P(\gs)$ to denote the partitions of a set~$\gs$. For example, $P([3]) = P(\{1,2,3\}) = \{\{1,2,3\}, \{1|2,3\}, \{1,2|3\}, \{1,3|2\}, \{1|2|3\}\}$, where $[\ell]$ is a shorthand notation for the set of integers from 1 to $\ell$. 
In the equation above, $\pi$ is a partition of $\gs$ while $|\pi|$ is the number of parts in the partition. The number of partitions of $[\ell]$ is given by the $\ell-$th Bell number \cite{becker1948arithmetic}. The photon-number moments appearing in the previous equation are calculated using Eq.~\eqref{eq:moments_general}. 

For cumulants where each mode appears only once, one equivalently writes a recursion ~\cite{zhong2021phase},
\begin{align} \label{eq:cumulant_recursive}
	\langle\langle\nh_1 \dots \nh\l\rangle\rangle = \braket{\nh_{1} \dots \nh\l} - \sum_{\pi\in P'([\ell])}\prod_{\bs \in \pi}\langle\langle \prod_{j\in \bs} \nh_j  \rangle\rangle.
\end{align}
Here, $P'([\ell])$ is the set of partitions of $[\ell]$ excluding the universal partition $\{1,2,\ldots,\ell\}$. Thus we see that cumulants are moments from which we remove some lower order terms.

\begin{figure*}[!t]
	\captionsetup{justification = raggedright, singlelinecheck = false}
	\begin{center}
	\input{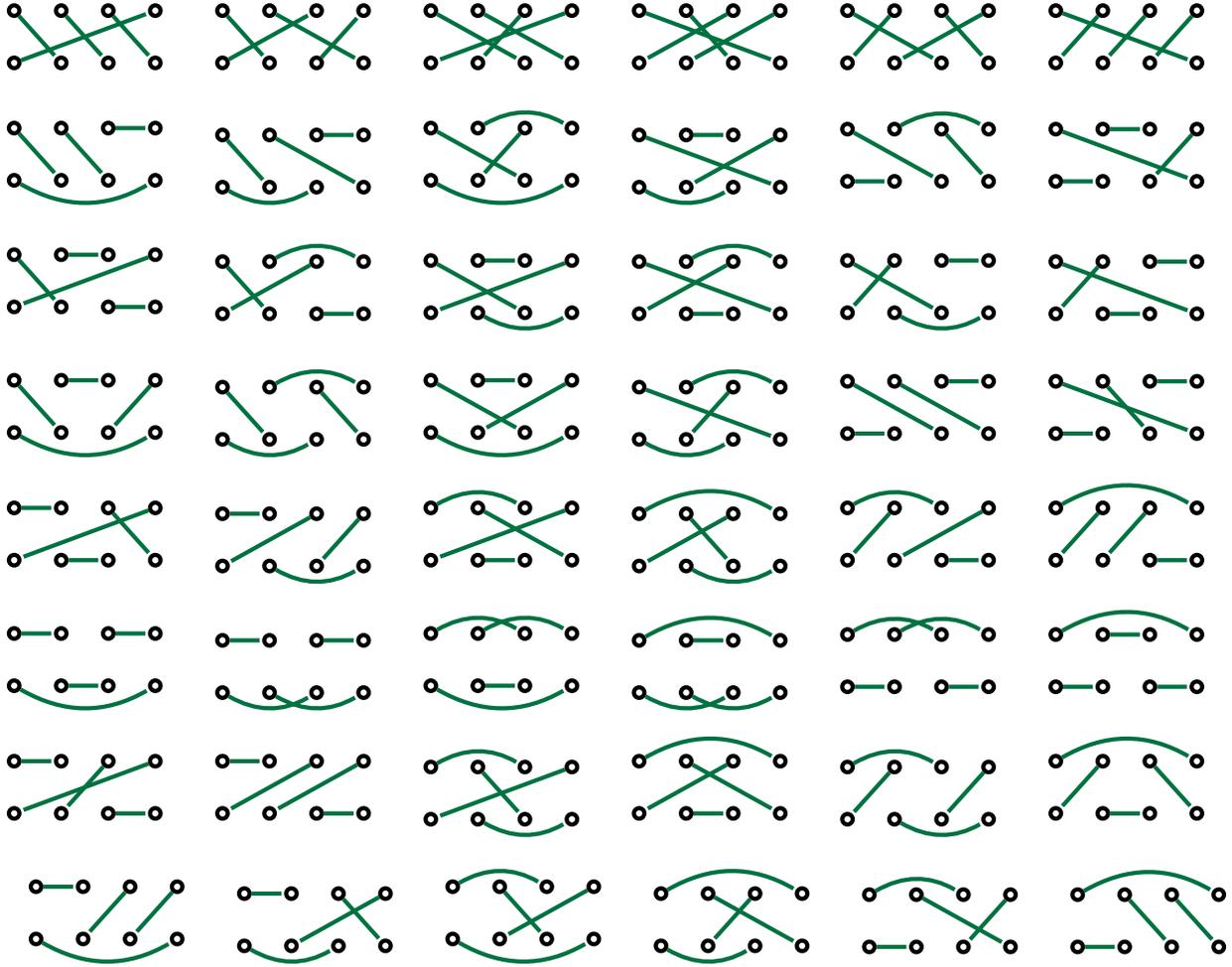}
	\end{center}
	\caption{\label{fig:lpm_4}Perfect matchings contributing to the fourth order cumulant  $\braket{\braket{\nh_1\nh_2\nh_3 \nh_4}}$ of a zero-displacement Gaussian state.}
	
\end{figure*}

Below we evaluate the first four cumulants (the last one only for zero-displacement) for Gaussian states, %
\begin{subequations}
	\begin{align}
		&\langle\langle\nh_1\rangle\rangle = \n_{11} + |\ab_1|^2, \label{eq:cumul1}\\
		&\langle\langle\nh_1 \nh_2\rangle\rangle= \label{eq:cumul2}\\
		&\quad |\m_{1 2}|^2 + |\n_{1 2}|^2 + (\ab_1\ab_2\m_{1 2}^* + \ab_1\ab_2^*\n_{1 2} + \text{c.c.}), \nonumber\\
		&\langle\langle\nh_1 \nh_2 \nh_3\rangle\rangle =  \label{eq:cumul3}  \\
		&\m_{1 3}^*\m_{1 2}\n_{2 3} + \m_{2 3}^*\m_{1 2}\n_{1 3} + \m_{2 3}^*\m_{1 3}\n_{1 2}  + \n_{1 3}^*\n_{1 2}\n_{2 3} \nonumber\\
		&\quad  + \ab_1\ab_2\m_{1 3}^*\n_{2 3} + \ab_1\ab_2\m_{2 3}^*\n_{1 3} + \ab_1\ab_3\m_{1 2}^*\n_{2 3}^*\nonumber\\
		&\quad + \ab_1\ab_3\m_{2 3}^*\n_{1 2} + \ab_1\ab_2^*\m_{1 3}^*\m_{2 3} + \ab_1\ab_2^*\n_{2 3}^*\n_{1 3} \nonumber\\
		&\quad + \ab_1\ab_3^*\m_{1 2}^*\m_{2 3} + \ab_1\ab_3^*\n_{1 2}\n_{2 3} + \ab_2\ab_3\m_{1 2}^*\n_{1 3}^* \nonumber\\
		&\quad + \ab_2\ab_3\m_{1 3}^*\n_{1 2}^* + \ab_2\ab_3^*\m_{1 2}^*\m_{1 3} + \ab_2\ab_3^*\n_{1 2}^*\n_{1 3}  \nonumber\\
		&\quad +\text{c.c.}\nonumber \\
		&\left.\langle\langle\nh_1 \nh_2 \nh_3 \nh_4\rangle\rangle\right|_{\zb = \bm{0}} =  \label{eq:cumul4}  \\
		&\ M_{13}^*M_{24}^*M_{12}M_{34} + M_{14}^*M_{23}^*M_{13}M_{24} + M_{14}^*M_{23}^*M_{12}M_{34} \nonumber \\
		&+ M_{13}^*N_{34}^*M_{12}N_{24} + M_{14}^*N_{23}^*M_{13}N_{24} + M_{14}^*M_{12}N_{23}N_{34} \nonumber \\
		&+ M_{23}^*N_{24}^*M_{13}N_{14} + M_{23}^*N_{24}^*M_{14}N_{13} + M_{23}^*N_{34}^*M_{12}N_{14} \nonumber \\
		&+ M_{23}^*N_{34}^*M_{14}N_{12} + M_{24}^*N_{13}^*M_{23}N_{14} + M_{24}^*N_{23}^*M_{13}N_{14} \nonumber \\
		&+ M_{24}^*N_{23}^*M_{14}N_{13} + M_{24}^*M_{12}N_{13}N_{34} + M_{24}^*M_{13}N_{12}N_{34} \nonumber \\
		&+ M_{34}^*N_{12}^*M_{23}N_{14} + M_{34}^*N_{12}^*M_{24}N_{13} + M_{34}^*M_{12}N_{13}N_{24} \nonumber \\
		&+ M_{34}^*M_{12}N_{14}N_{23} + M_{34}^*M_{13}N_{12}N_{24} + M_{34}^*M_{14}N_{12}N_{23} \nonumber \\
		&+ N_{13}^*N_{34}^*N_{12}N_{24} + N_{14}^*N_{23}^*N_{13}N_{24} + N_{14}^*N_{12}N_{23}N_{34}\nonumber \\
		&+\text{c.c.}   \nonumber
	\end{align}
\end{subequations}
More generally, using the recursion Eq.~\eqref{eq:cumulant_recursive}, it is readily seen that the cumulants are given by restricted subsets of perfect matchings. Consider first the case of $\ell=2$ modes. Eq.~\eqref{eq:cumulant_recursive} states that one should first calculate the loop Hafnian of a $2\ell \times 2\ell$ matrix ($2 \ell =4$) which in this case has $|\text{SPM}(2\ell = 4)| = 10$ terms. Then one must subtract the product of the terms corresponding to single-mode cumulants. The four single-pair matchings that are removed are coloured in red and purple in Fig.~\ref{fig:lpm_2} while the six single-pair matchings that survive and thus give the sought-after cumulant are coloured in green and blue and correspond with the six terms in Eq.~\eqref{eq:cumul2}. Note moreover that if we restrict to Gaussian states with zero displacement then the blue graphs also drop out leaving only the green perfect matchings.
This process can be repeated inductively to obtain the subset of single-pair matchings that describe a cumulant of a given order. For 3 modes, we use the same colour convention as described in the previous paragraph to separate the single-pair matching that are present in the third order cumulant in green and blue and the ones that drop out coloured in red and purple; the corresponding perfect matchings and monomials are shown in Fig.~\ref{fig:lpm_3} and Eq.~\eqref{eq:cumul3}. Finally, we show the loop-less perfect matchings contributing to the fourth order cumulant appearing in Fig.~\ref{fig:lpm_4} and Eq.~\eqref{eq:cumul4}. The colour scheme used for the different perfect matchings is further shown in Table~\ref{tab:color_legend}. This graph representation of cumulants highlights the adequacy of the connected correlation functions terminology.

\subsection{The Montrealer and Loop Montrealer}
It is desirable to have a deeper understanding of the properties of the perfect matchings that make it into the cumulants. To this end we introduce some graph-theoretic notation ~\cite{bjorklund2019faster}. Photon-number cumulants can be built using alternating walks. A walk is defined as a sequence of vertices $\bm{w} = \{w_0, w_1, \dots, w_n\}$ where no elements are allowed to be repeated except for the first $w_0$ and the last $w_n$. A walk $\bm{w} = \{w_0, w_1, \dots, w_n\}$ is said to be $Y$-alternating if and only if, $\forall\, i<n$,
\begin{itemize}
	\item $i$ is even: $(w_i, w_{i+1})\notin Y$,
	\item $i$ is odd: $(w_i, w_{i+1})\in Y$,
\end{itemize}
and moreover
\begin{itemize}
	\item If the walk has no loops then $\bm{w}$ has to be closed $(w_0 = w_n)$ .
	\item If the walk has loops then there are only two of them and they lie at the endpoints of the walk ($w_0 = w_1$ and $w_{n-1}=w_n$).
\end{itemize}
Since in our graph representation vertices $i \text{ and } i+\ell$ represent the same mode, we will be interested in the $Y$-alternating walks over the perfect matching
\begin{align}\label{eq:Ydef}
	Y = \{(1, \ell+1), (2, \ell+2), \dots,(\ell, 2\ell)\},
\end{align} 
represented graphically in Fig.~\ref{fig:XY}(A). The $Y$-alternating walks for the $Y$ defined above will be of length $n= 2\ell$ if there are no loops and of length $n+1= 2 \ell + 1$ if a pair of loops is present.

\begin{figure}[!t]
	\captionsetup{justification = raggedright, singlelinecheck = false}
	\begin{tikzpicture}[scale=.7, shorten >=1pt, auto, node distance=5cm, ultra thick]
    \tikzstyle{node_style} = [circle,draw=black, inner sep=0pt, minimum size=4pt]
    \tikzstyle{edge_style} = [-,draw=mtl_yellow, line width=2, ultra thick]
    \tikzstyle{edge_styleg} = [-,draw=mtl_red , line width=2, ultra thick]
    \tikzstyle{edge_stylegg} = [-,draw=white , line width=2, ultra thick]
    \tikzstyle{text_style} = [circle, inner sep=0pt, minimum size=12pt, font = \normalsize]

    \node[node_style] (v1) at (0,1) {};
    \node[node_style] (v2) at (0.9,1) {};
    \node[node_style] (v3) at (1.8,1) {};
    \node[node_style] (v4) at (2.7,1) {};
    \node[node_style] (v5) at (0,0) {};
    \node[node_style] (v6) at (0.9,0) {};
    \node[node_style] (v7) at (1.8,0) {};
    \node[node_style] (v8) at (2.7,0) {};

    \draw[edge_stylegg]  (v5) to [loop right, in=-30,out=-150,looseness=10] (v5);
    \draw[edge_stylegg]  (v1) to [loop right, in=30,out=150,looseness=10] (v1);
    \draw[edge_stylegg]  (v8) to [loop right, in=-30,out=-150,looseness=10] (v8);
    \draw[edge_stylegg]  (v4) to [loop right, in=30,out=150,looseness=10] (v4);
    
    \draw[edge_styleg]  (v1) edge (v6);
    \draw[edge_styleg]  (v2) edge (v5);
    \draw[edge_styleg]  (v3) edge (v8);
    \draw[edge_styleg]  (v4) edge (v7);
    \draw[edge_stylegg]  (v1) edge [bend left] node[below] {} (v4);
    \draw[edge_stylegg]  (v5) edge [bend right] node[below] {} (v8);

    \draw[edge_style]  (v1) edge (v5);
    \draw[edge_style]  (v2) edge (v6);
    \draw[edge_style]  (v3) edge (v7);
    \draw[edge_style]  (v4) edge (v8);

    \node[text_style] at (1.5,-1) {$\subseteq\langle\langle\hat{n}_1\hat{n}_2\rangle\rangle \times\langle\langle\hat{n}_3\hat{n}_4\rangle\rangle$};
    
 \end{tikzpicture} \vspace{-.5 cm}
\
$\quad$
\begin{tikzpicture}[scale=.7, shorten >=1pt, auto, node distance=1cm, ultra thick]
    \tikzstyle{node_style} = [circle,draw=black, inner sep=0pt, minimum size=4pt]
    \tikzstyle{edge_style} = [-,draw=mtl_yellow, line width=2, ultra thick]
    \tikzstyle{edge_styleg} = [-,draw=mtl_red , line width=2, ultra thick]
    \tikzstyle{edge_stylegg} = [-,draw=white , line width=2, ultra thick]
    \tikzstyle{text_style} = [circle, inner sep=0pt, minimum size=12pt, font = \normalsize]

    \node[node_style] (v1) at (0,1) {};
    \node[node_style] (v2) at (0.9,1) {};
    \node[node_style] (v3) at (1.8,1) {};
    \node[node_style] (v4) at (2.7,1) {};
    \node[node_style] (v5) at (0,0) {};
    \node[node_style] (v6) at (0.9,0) {};
    \node[node_style] (v7) at (1.8,0) {};
    \node[node_style] (v8) at (2.7,0) {};

    \draw[edge_stylegg]  (v5) to [loop right, in=-30,out=-150,looseness=10] (v5);
    \draw[edge_stylegg]  (v1) to [loop right, in=30,out=150,looseness=10] (v1);
    \draw[edge_stylegg]  (v8) to [loop right, in=-30,out=-150,looseness=10] (v8);
    \draw[edge_stylegg]  (v4) to [loop right, in=30,out=150,looseness=10] (v4);

    \draw[edge_styleg]  (v2) edge (v7);
    \draw[edge_styleg]  (v3) edge (v6);
    \draw[edge_styleg]  (v1) edge [bend left] node[below] {} (v4);
    \draw[edge_styleg]  (v5) edge [bend right] node[below] {} (v8);
    
    \draw[edge_style]  (v1) edge (v5);
    \draw[edge_style]  (v2) edge (v6);
    \draw[edge_style]  (v3) edge (v7);
    \draw[edge_style]  (v4) edge (v8);

    \node[text_style] at (1.5,-1) {$\subseteq\langle\langle\hat{n}_1\hat{n}_4\rangle\rangle \times \langle\langle\hat{n}_2\hat{n}_3\rangle\rangle$};
    
\end{tikzpicture} \vspace{-.5 cm}

\begin{tikzpicture}[scale=.7, shorten >=1pt, auto, node distance=5cm, ultra thick]
    \tikzstyle{node_style} = [circle,draw=black, inner sep=0pt, minimum size=4pt]
    \tikzstyle{edge_style} = [-,draw=mtl_yellow, line width=2, ultra thick]
    \tikzstyle{edge_styleg} = [-,draw=mtl_purple , line width=2, ultra thick]
    \tikzstyle{edge_stylegg} = [-,draw=white , line width=2, ultra thick]
    \tikzstyle{text_style} = [circle, inner sep=0pt, minimum size=12pt, font = \normalsize]

    \node[node_style] (v1) at (0,1) {};
    \node[node_style] (v2) at (0.9,1) {};
    \node[node_style] (v3) at (1.8,1) {};
    \node[node_style] (v4) at (2.7,1) {};
    \node[node_style] (v5) at (0,0) {};
    \node[node_style] (v6) at (0.9,0) {};
    \node[node_style] (v7) at (1.8,0) {};
    \node[node_style] (v8) at (2.7,0) {};

    \draw[edge_stylegg]  (v5) to [loop right, in=-30,out=-150,looseness=10] (v5);
    \draw[edge_stylegg]  (v1) to [loop right, in=30,out=150,looseness=10] (v1);
    \draw[edge_stylegg]  (v5) edge [bend right] node[below] {} (v8);
    
    \draw[edge_styleg]  (v5) edge (v3);
    \draw[edge_styleg]  (v1) edge (v6);
    \draw[edge_styleg]  (v2) edge (v7);
    \draw[edge_styleg]  (v8) to [loop right, in=-30,out=-150,looseness=10] (v8);
    \draw[edge_styleg]  (v4) to [loop right, in=30,out=150,looseness=10] (v4);

    \draw[edge_style]  (v1) edge (v5);
    \draw[edge_style]  (v2) edge (v6);
    \draw[edge_style]  (v3) edge (v7);
    \draw[edge_style]  (v4) edge (v8);
    
    \node[text_style] at (1.5,-1){$\subseteq\langle\langle\hat{n}_1\hat{n}_2\hat{n}_3\rangle\rangle \times \langle\langle\hat{n}_4\rangle\rangle$};
    
 \end{tikzpicture} \vspace{-.5 cm}
\
$\quad$
\begin{tikzpicture}[scale=.7, shorten >=1pt, auto, node distance=5cm, ultra thick]
    \tikzstyle{node_style} = [circle,draw=black, inner sep=0pt, minimum size=4pt]
    \tikzstyle{edge_style} = [-,draw=mtl_yellow, line width=2, ultra thick]
    \tikzstyle{edge_styleg} = [-,draw=mtl_purple , line width=2, ultra thick]
    \tikzstyle{edge_stylegg} = [-,draw=white , line width=2, ultra thick]
    \tikzstyle{text_style} = [circle, inner sep=0pt, minimum size=12pt, font = \normalsize]

    \node[node_style] (v1) at (0,1) {};
    \node[node_style] (v2) at (0.9,1) {};
    \node[node_style] (v3) at (1.8,1) {};
    \node[node_style] (v4) at (2.7,1) {};
    \node[node_style] (v5) at (0,0) {};
    \node[node_style] (v6) at (0.9,0) {};
    \node[node_style] (v7) at (1.8,0) {};
    \node[node_style] (v8) at (2.7,0) {};

    \draw[edge_stylegg]  (v5) to [loop right, in=-30,out=-150,looseness=10] (v5);
    \draw[edge_stylegg]  (v8) to [loop right, in=-30,out=-150,looseness=10] (v8);
    
    \draw[edge_styleg]  (v2) edge (v7);
    \draw[edge_styleg]  (v3) edge (v6);
    \draw[edge_styleg]  (v4) to [loop right, in=30,out=150,looseness=10] (v4);
    \draw[edge_styleg]  (v1) to [loop right, in=30,out=150,looseness=10] (v1);
    \draw[edge_styleg]  (v5) edge [bend right] node[below] {} (v8);

    \draw[edge_style]  (v1) edge (v5);
    \draw[edge_style]  (v2) edge (v6);
    \draw[edge_style]  (v3) edge (v7);
    \draw[edge_style]  (v4) edge (v8);

    \node[text_style] at (1.5,-1){$\subseteq\langle\langle\hat{n}_1\hat{n}_4\rangle\rangle \times \langle\langle\hat{n}_2\hat{n}_3\rangle\rangle$};
    
 \end{tikzpicture} \vspace{-.5 cm}
	\caption{\label{fig:disjoint} Examples of graphs appearing in a fourth order moment that are not $Y$-alternating walks. Such disjoint graphs can always be attributed to lower order cumulants
		not contributing to the fourth order cumulant as shown by Eq.~\eqref{eq:cumulant_recursive}.}
\end{figure}
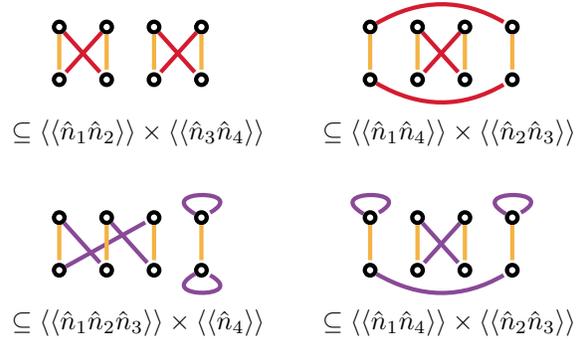

Note that every perfect matching contributing to a cumulant will make, when joined with $Y$, a $Y$-alternating walk, and reciprocally, that any perfect matching not contributing to the cumulant will be unable to make a $Y-$alternating walk as illustrated graphically in Fig.~\ref{fig:XY}.
Indeed, if the union of an element from $X\in \text{PMP}$ and $Y$ does not make a $Y$-alternating then it must be that the term corresponding to $X$ in a photon-number moment can be obtained as the product of two lower-order cumulants; however, the definition in Eq.~\eqref{eq:cumulant_recursive} removes these product terms from a given photon-number moment to form a photon-number cumulant. Examples of such excluded graphs are given in Fig.~\ref{fig:disjoint}.

For a given graph with $2\ell$ vertices we define the set $\text{RPMP}(2\ell)$ (restricted perfect matching permutations) as the set of perfect matchings without loops that when interleaved with $Y$ give a $Y$-alternating walk.
Similarly, we define the set $\text{RSPM}(2\ell)$ (restricted single pair matchings) as the set of perfect matchings including loops that when interleaved with $Y$ give a $Y$-alternating walk.
These two sets capture the perfect matchings that contribute to the photon-number cumulants of a Gaussian state. We then introduce the Montrealer and the loop Montrealer of a $2\ell\times2 \ell$ matrix $\A$ as
\begin{align}
	\mtl \A = \sum_{\bs\,\in\,\text{RPMP}(2\ell)}\prod_{(i,j)\,\in\,\bs}A_{ij}\,,\\
	\lmtl \A = \sum_{\bs\,\in\,\text{RSPM}(2\ell)}\prod_{(i,j)\,\in\,\bs}A_{ij}\,.
\end{align}
For zero-displacement Gaussian states, the Montrealer ($\mtl$) of the adjacency matrix $\A$ coincides with the $\ell-$th order photon-number cumulant in Eq.~\eqref{eq:cumulant_recursive}; %
correspondingly the loop Montrealer of $\fdiag(\A,\zb^*)$ gives the photon-number cumulants of the Gaussian state with finite displacement $\bm{\zeta}$ and adjacency matrix $\A$.

\begin{figure}[!t]
	\centering
	\captionsetup{justification = raggedright, singlelinecheck = false}
		\begin{tikzpicture}[scale=.7, shorten >=1pt, auto, node distance=1cm, ultra thick]
		\tikzstyle{node_style} = [circle,draw=black, inner sep=0pt, minimum size=4pt]
		\tikzstyle{node_blank} = [circle,draw=white, inner sep=0pt, minimum size=4pt]
		\tikzstyle{edge_blank} = [-,draw=white, line width=2, thick, dashed]
		\tikzstyle{edge_styleg1} = [-,draw=mtl_yellow , line width=2, ultra thick]
		\tikzstyle{edge_styleg2} = [-,draw=mtl_green , line width=2, ultra thick]
		\tikzstyle{text_style} = [circle, inner sep=0pt, minimum size=14pt, font = \large]
		\tikzstyle{text_style2} = [circle, inner sep=0pt, minimum size=14pt, font = \Large]
		
		\node[node_style] (v1) at (0,2.5) {};
		\node[node_style] (v2) at (1.75,2.5) {};
		\node[node_style] (v3) at (3.5,2.5) {};
		\node[node_blank] (v4) at (5.25,2.5) {};
		\node[node_style] (v5) at (7,2.5) {};
		\node[node_style] (v6) at (8.75,2.5) {};
		\node[node_style] (v7) at (0,0) {};
		\node[node_style] (v8) at (1.75,0) {};
		\node[node_style] (v9) at (3.5,0) {};
		\node[node_blank] (v10) at (5.25,0) {};
		\node[node_style] (v11) at (7,0) {};
		\node[node_style] (v12) at (8.75,0) {};
		\node[node_blank] (xx) at (4.5,1.5) {};
		\node[node_blank] (yy) at (6,1) {};
		
		\draw[edge_styleg1]  (v1) edge (v7);
		\draw[edge_styleg1]  (v2) edge (v8);
		\draw[edge_styleg1]  (v3) edge (v9);
		\draw[edge_styleg1]  (v5) edge (v11);
		\draw[edge_styleg1]  (v6) edge (v12);
		
		\node[text_style] at (0,3) {$1$};
		\node[text_style] at (1.75,3) {$2$};
		\node[text_style] at (3.5,3) {$3$};
		\node[text_style] at (7,3) {$\ell-1$};
		\node[text_style] at (8.75,3) {$\ell$};
		\node[text_style] at (0,-.5) {$\ell+1$};
		\node[text_style] at (1.75,-.5) {$\ell+2$};
		\node[text_style] at (3.5,-.5) {$\ell+3$};
		\node[text_style] at (7,-.5) {$2\ell-1$};
		\node[text_style] at (8.75,-.5) {$2\ell$};
		
		\node[text_style2] at (5.25,0.5) {$\dots$};
		\node[text_style2] at (-2,1.5) {$\qquad\quad Y :$};
		
	\end{tikzpicture} \vspace{0 cm}\hspace{-0.3 cm}
	
	\begin{tikzpicture}[scale=.7, shorten >=1pt, auto, node distance=1cm, ultra thick]
		\tikzstyle{node_style} = [circle,draw=black, inner sep=0pt, minimum size=4pt]
		\tikzstyle{node_blank} = [circle,draw=white, inner sep=0pt, minimum size=4pt]
		\tikzstyle{edge_blank} = [-,draw=white, line width=2, thick, dashed]
		\tikzstyle{edge_styleg1} = [-,draw=mtl_purple , line width=2, ultra thick]
		\tikzstyle{edge_styleg2} = [-,draw=mtl_green , line width=2, ultra thick]
		\tikzstyle{text_style} = [circle, inner sep=0pt, minimum size=14pt, font = \large]
		\tikzstyle{text_style2} = [circle, inner sep=0pt, minimum size=14pt, font = \Large]
		
		\node[node_style] (v1) at (0,2.5) {};
		\node[node_style] (v2) at (1.75,2.5) {};
		\node[node_style] (v3) at (3.5,2.5) {};
		\node[node_blank] (v4) at (5.25,2.5) {};
		\node[node_style] (v5) at (7,2.5) {};
		\node[node_style] (v6) at (8.75,2.5) {};
		\node[node_style] (v7) at (0,0) {};
		\node[node_style] (v8) at (1.75,0) {};
		\node[node_style] (v9) at (3.5,0) {};
		\node[node_blank] (v10) at (5.25,0) {};
		\node[node_style] (v11) at (7,0) {};
		\node[node_style] (v12) at (8.75,0) {};
		\node[node_blank] (xx) at (4.3,1.45) {};
		\node[node_blank] (yy) at (6.2,1.05) {};
		
		\draw[edge_styleg2]  (v1) edge (v8);
		\draw[edge_styleg2]  (v2) edge (v9);
		\draw[edge_styleg2]  (v5) edge (v12);
		\draw[edge_styleg2]  (v6) edge (v7);
		
		\draw[edge_styleg2]  (v3) edge (xx);
		\draw[edge_styleg2]  (v11) edge (yy);
		
		\node[text_style] at (0,3) {$1$};
		\node[text_style] at (1.75,3) {$2$};
		\node[text_style] at (3.5,3) {$3$};
		\node[text_style] at (7,3) {$\ell-1$};
		\node[text_style] at (8.75,3) {$\ell$};
		\node[text_style] at (0,-.5) {$\ell+1$};
		\node[text_style] at (1.75,-.5) {$\ell+2$};
		\node[text_style] at (3.5,-.5) {$\ell+3$};
		\node[text_style] at (7,-.5) {$2\ell-1$};
		\node[text_style] at (8.75,-.5) {$2\ell$};
		
		\node[text_style2] at (5.25,0.5) {$\dots$};
		\node[text_style2] at (-2,1.5) {$\qquad X_{\text{fid}} :$};
		
	\end{tikzpicture} \vspace{0 cm}\hspace{-0.3 cm}
	
	\begin{tikzpicture}[scale=.7, shorten >=1pt, auto, node distance=1cm, ultra thick]
		\tikzstyle{node_style} = [circle,draw=black, inner sep=0pt, minimum size=4pt]
		\tikzstyle{node_blank} = [circle,draw=white, inner sep=0pt, minimum size=4pt]
		\tikzstyle{edge_blank} = [-,draw=white, line width=2, thick, dashed]
		\tikzstyle{edge_styleg1} = [-,draw=mtl_yellow , line width=2, ultra thick]
		\tikzstyle{edge_styleg2} = [-,draw=mtl_green , line width=2, ultra thick]
		\tikzstyle{text_style} = [circle, inner sep=0pt, minimum size=14pt, font = \large]
		\tikzstyle{text_style2} = [circle, inner sep=0pt, minimum size=14pt, font = \Large]
		
		\node[node_style] (v1) at (0,2.5) {};
		\node[node_style] (v2) at (1.75,2.5) {};
		\node[node_style] (v3) at (3.5,2.5) {};
		\node[node_blank] (v4) at (5.25,2.5) {};
		\node[node_style] (v5) at (7,2.5) {};
		\node[node_style] (v6) at (8.75,2.5) {};
		\node[node_style] (v7) at (0,0) {};
		\node[node_style] (v8) at (1.75,0) {};
		\node[node_style] (v9) at (3.5,0) {};
		\node[node_blank] (v10) at (5.25,0) {};
		\node[node_style] (v11) at (7,0) {};
		\node[node_style] (v12) at (8.75,0) {};
		\node[node_blank] (xx) at (4.3,1.45) {};
		\node[node_blank] (yy) at (6.2,1.05) {};
		
		\draw[edge_styleg1]  (v1) edge (v7);
		\draw[edge_styleg1]  (v2) edge (v8);
		\draw[edge_styleg1]  (v3) edge (v9);
		\draw[edge_styleg1]  (v5) edge (v11);
		\draw[edge_styleg1]  (v6) edge (v12);
		
		\draw[edge_styleg2]  (v1) edge (v8);
		\draw[edge_styleg2]  (v2) edge (v9);
		\draw[edge_styleg2]  (v5) edge (v12);
		\draw[edge_styleg2]  (v6) edge (v7);
		
		\draw[edge_styleg2]  (v3) edge (xx);
		\draw[edge_styleg2]  (v11) edge (yy);
		
		\node[text_style] at (0,3) {$1$};
		\node[text_style] at (1.75,3) {$2$};
		\node[text_style] at (3.5,3) {$3$};
		\node[text_style] at (7,3) {$\ell-1$};
		\node[text_style] at (8.75,3) {$\ell$};
		\node[text_style] at (0,-.5) {$\ell+1$};
		\node[text_style] at (1.75,-.5) {$\ell+2$};
		\node[text_style] at (3.5,-.5) {$\ell+3$};
		\node[text_style] at (7,-.5) {$2\ell-1$};
		\node[text_style] at (8.75,-.5) {$2\ell$};
		
		\node[text_style2] at (5.25,0.5) {$\dots$};
		\node[text_style2] at (-2,1.5) {$X_{\text{fid}}\cup Y :$};
		
	\end{tikzpicture} \vspace{0 cm}\hspace{-0.3 cm}
	
	\caption{Graphical representation of the perfect matching $Y$ defining the alternating walks, the fiducial restricted perfects matching $X_\text{fid}$ and their union. 
		\label{fig:XY}}
\end{figure}

The set RPMP can be constructed as follows: we first need a fiducial element from which all others will be derived by appropriate transformation. A convenient fiducial element for arbitrary $\ell$ is given by
\begin{align}
	X_{\text{fid}} = \{ (1,\ell+2),(2,\ell+3),\dots,(\ell-1,2\ell),(\ell,\ell+1)\}.
\end{align}
Note that if we take the edges in $X_{\text{fid}}$ together with the edges in $Y$ we obtain a $Y$-alternating walk of length precisely $2\ell$. 

We now note that the set of permutations that leave invariant the set perfect matching defining the alternating walk, $Y$ in Eq.~\eqref{eq:Ydef} are of two types
\begin{enumerate}
	\item Products of transpositions that permute vertices $i$ and $i+\ell$ for $1 \leq i \leq \ell$. There are $2^\ell$ of these products. 
	\item Any permutation of $\ell$ objects that jointly permutes the sets $\{1,2,\ldots,\ell\}$ and $\{\ell+1,\ell+2,\ldots,2\ell\}$. There are a total of $\ell!$ permutations of $\ell$ objects. 
\end{enumerate}
Thus by applying the transformations described in the list above on $X_{\text{fid}}$, one should be able to obtain all the elements in $\text{RPMP}(2\ell)$. However, if we apply all the $2^\ell \ell!  = (2 \ell)!!$ permutation on $X_{\text{fid}}$ we will generate each element of $\text{RPMP}(2\ell)$ a total of $2\ell$ times. 

It is desirable to restrict the permutations so that each element is generated only once, thus we use the following set of permutation on $X_{\text{fid}}$ to generate the set $\text{RPMP}(2\ell)$
\begin{enumerate}
	\item Products of transpositions that permute vertices $i$ and $i+\ell$ for $2 \leq i \leq \ell$. There are $2^{\ell-1}$ of these products. 
	\item Any permutation of $\ell$ objects that jointly permutes the sets $\{2,\ldots,\ell\}$ and $\{\ell+2,\ldots,2\ell\}$. There are a total of $(\ell-1)!$ permutations of $\ell-1$ objects. 
\end{enumerate}
Note that the above construction gives $|\text{RPMP}(2\ell)| = (2 \ell -2)!!$.

To obtain the perfect matchings RSPM of the loop Montrealer we use the same algorithm as the one we previously defined for the RPMP but with an additional step. All the perfect matchings of the Montrealer are also included in the loop Montrealer, but we now add new perfect matchings built by taking each element for RPMP and breaking a single edge from all of them, turning it into two individual loops. We end up breaking all edges from the previous graphs but always only one at the time. Hence, the additional graphs of the loop Montrealer containing single-pair perfect matchings will always contain precisely two loops. Since all graphs of the Montrealer have exactly $\ell$ edges we find that $|\text{RSPM}(2\ell)| = (\ell+1)(2\ell-2)!!$ .

\subsection{Properties of the photon-number cumulants and the Montrealer}
The photon-number cumulants satisfy a number of interesting identities, which also translate into identities for the Montrealer.  If we let $\bm{\Lambda} = \oplus_{i=1}^\ell \lambda_i$ be a complex diagonal matrix, and $\bm{A} = \bm{A}^T \in \mathbb{C}^{2\ell \times 2\ell}$ be an arbitrary complex symmetric matrix then
\begin{align}
	\mtl \left[\left(\bm{\Lambda}\oplus\bm{\Lambda}^*\right)\A\left(\bm{\Lambda}\oplus\bm{\Lambda}^*\right)\right] &= \left(\prod_{i=1}^\ell |\lambda_i|^2 \right) \mtl \A.
\end{align}
Note that the phases of the $\lambda_i$ do not affect the value of the final result, which is only affected by the product of the magnitudes. This first observation about the phases physically corresponds to the fact that applying local rotations to the modes right before measurement, i.e., letting
\begin{align}
	\ah_j \to \ah_j e^{i \theta_j},
\end{align}
does not modify the photon-number statistics. 
The second observation is simply related to the fact that a product of photon-number operators  of different modes,
\begin{align}
	\braket{\nh_1 \nh_2 \ldots \nh_r} = \braket{\ah_1^\dagger \ah_2^\dagger \ldots \ah_r^\dagger \ah_1 \ah_2 \ldots \ah_r},
\end{align}
is trivially normal ordered and thus under loss channels with energy transmission $\eta_i = |\lambda_i|^2$ will transform precisely as in the last equation.

Note that since the Montrealer is constructed around the idea of invariance with respect to the action of the same permutation on vertices $(1,\ldots,\ell)$ and $(\ell+1,\ldots,2\ell)$ then it must hold that
\begin{align}
	\mtl \left(\bm{P}\oplus\bm{P}\right)\A\left(\bm{P}^T\oplus\bm{P}^T\right) &= \mtl\A, %
\end{align}
where $\bm{P}$ is a permutation matrix. Physically, this equation comes from the fact that a cumulant is invariant under a permutation of the random variables given as its argument. This permutation invariance property is also shared by other matrix functions such as the Determinant, the Permanent, the Hafnian and the Pfaffian.

Since the Montrealer describes the genuine correlations of a set of modes, it must hold that if the four $\ell \times \ell$ blocks of $\bm{A}$ have direct sum structure, then the Montrealer of the associated adjacency matrix must be zero %
\begin{align}
	\mtl \left(\begin{array}{cc|cc}
		\bm{M}_1 & 0 & \bm{N}_1 & 0 \\
		0 & \bm{M}_2 & 0 & \bm{N}_2  \\
		\hline
		\bm{N}_1^T & 0 & \bm{M}_3 & 0 \\
		0 & \bm{N}_2^T & 0 & \bm{M}_4  \\
	\end{array} \right) = 0,
\end{align}
where $\bm{M}_1$, $\bm{M}_3$, $\bm{N}_1$ have dimensions $k \times k$ and $\bm{M}_2$, $\bm{M}_4$, $\bm{N}_2$ have dimensions $\ell - k \times \ell - k$ and $\bm{M}_i = \bm{M}_i^T$.
Physically, this comes from the fact that if the input matrix has the block-diagonal form shown above, then clearly there are no global correlations between all modes, but only between the first $k$ modes and separately between the last $\ell - k$ modes.
This property of the Montrealer is not shared by the Permanent, the Determinant, the Hafnian or the Pfaffian which distribute as a product when acted over a direct sum.

Finally, note that if we write the argument of the Montrealer in block form
\begin{align}
	\mtl \left( \begin{array}{c|c}
		\bm{M}^* & \bm{N} \\
		\hline 
		\bm{N}^T & \bm{M} 
	\end{array} \right),
\end{align}
then the final value is independent of the diagonal elements of $\bm{M}$ and $\bm{N}$ if the number of modes is greater than 1. Note that in general, the diagonal elements of the argument do not contribute, thus we need not worry about diagonal values of the $\bm{M}$. As for the diagonal elements of $\bm{N}$ they appear in perfect matchings that do not make an $Y$-alternating walk and thus cannot contribute to a cumulant.

In the context of Gaussian states this will imply that if $\bm{N} = \oplus_{i=1}^\ell n_i$ then the photon-number cumulant of the Gaussian state will be given by 
\begin{subequations}
	\begin{align}
		\braket{\braket{\nh_1 \ldots \nh_\ell}} =& \mtl \left( \begin{array}{c|c}
			\bm{M}^* & \oplus_{i=1}^\ell n_i \\
			\hline 
			\oplus_{i=1}^\ell n_i& \bm{M}
		\end{array} \right) \\
		=& \mtl \left( \begin{array}{c|c}
			\bm{M}^* &0 \\
			\hline 
			0& \bm{M} 
		\end{array} \right),
	\end{align}
\end{subequations}
and thus will be independent of the mean-photon numbers $n_i = N_{ii}$ (for $\ell \geq 2$).
Moreover, note that in this case if $\ell$ is odd then the photon-number cumulants will be zero. This is because the weight associated to the perfect matchings linking the bottom and lower halves of the graph is zero, and for an odd number of columns one can not build a $Y$-alternating walk using only the perfect matchings corresponding to $\M$ which are the perfect matching that do not cross between the two halves of the graph.
For even $\ell$ the photon-number cumulants will in general be nonzero as one can perfect match both the top and lower half of the graph separately.

\subsection{Complexity theory: Connection with Hamiltonian walks}
In this section we establish a connection between the photon-number cumulants of Gaussian states, the Montrealer matrix function and a well-known problem in complexity theory: counting the number of Hamiltonian cycles on a graph.
Given a graph $G = (V,E)$ where $V$ is a set of vertices and $E$ a set of edges connecting vertices, we define a Hamiltonian cycle (or circuit) as a path that visits each vertex of the graph only once using the edges of the graph and in which the first and last vertex of the path coincide. Given a graph, counting the number of Hamiltonian cycles it has is a $\#P-$complete problem as shown by Valiant~\cite{valiant1979complexity}.

Just like counting perfect matching on a (bipartite) graph can be associated with the (Permanent) Hafnian of its adjacency matrix, one can introduce a matrix function that calculates the number of Hamiltonian cycles given the adjacency matrix of the graph. Thus the Hamiltonian cycle polynomial ($\ham$)~\cite{fefferman2014power} is defined as
\begin{align}
	\ham  \bm{B} = \sum_{\sigma \in H_n} \prod_{i=1}^\ell B_{i,\sigma(i)},
\end{align}
where $\bm{B}$ is the adjacency matrix of the graph, $H_n\,(|H_n| = (n-1)!)$ is the set of permutations of $n$ objects having exactly one cycle (equivalently they are all the $n$-cycle permutations). Note that at this point $\bm{B}$ need not be a $(0,1)$-matrix and can in general be defined over the complex numbers.

\begin{figure*}[!t]
	\captionsetup{justification = raggedright, singlelinecheck = false}
	\begin{center}
	\input{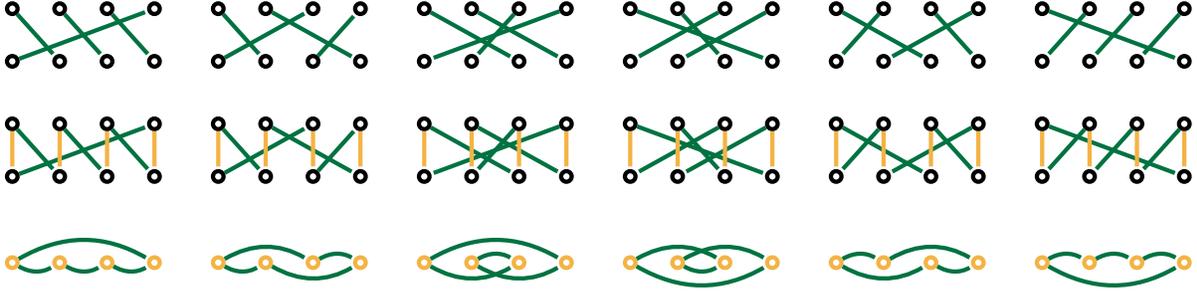}
	\end{center}
	\caption{\label{fig:merging} Turning $Y$-alternating walks on a graph with $2\ell = 8$ vertices into Hamiltonian cycles on a graph with $\ell=4$ vertices.}
\end{figure*}

With these definitions, we can now show that
\begin{align}\label{eq:mtlham}
	\mtl \left( \begin{array}{c|c}
		0_n & \bm{B} \\
		\hline
		\bm{B}^T & 0_n
	\end{array} \right) = \ham \bm{B}.
\end{align}
To prove this identity, first recall that the diagonal blocks of the matrices entering the Montrealer represent edges linking vertices within the same upper or lower half of the graph (cf. Fig.~\ref{fig:graph_legend}). In the equation above these types of edges have zero weights and thus we are allowing for alternating walks over $Y$ that necessarily also alternate between the upper and lower half of the graph. Thus we can without loss of generality merge the upper and lower half, literally merging vertices $i$ and $i+\ell$ of the graph and thus it holds that an alternating walk over $Y$ must be also a cycle over the graph $(1 \leftrightarrow \ell+1,2 \leftrightarrow \ell+2,\ldots,\ell \leftrightarrow 2 \ell)$. This process is represented graphically in Fig.~\ref{fig:merging}. Eq.~\eqref{eq:mtlham} implies that in general calculating photon-number cumulants of Gaussian states is a $\#P-$hard problem. This is the case because given an arbitrary $(0,1)$ adjancency matrix of an undirected graph $\tilde{\bm{B}} = \tilde{\bm{B}}^T$ we can construct a thermal state with $\bm{N} = \bm{N}^T = \lambda \I_n + \tilde{\bm{B}}$  and $\bm{M} = 0$ and $\lambda$ chosen so that $\bm{N}$ is positive semidefinite. This Gaussian state will have cumulants given by $\mtl \left( \begin{smallmatrix}
	0_n & \bm{N} \\
	\bm{N}^T & 0_n
\end{smallmatrix} \right) = \ham \N$ and thus calculating photon-number cumulants is as hard as counting Hamiltonian cycles of undirected graphs.

While both the photon-number distribution of Gaussian states  and  the multinormal distribution are similar in that their moments are given by loop Hafnians, they sharply differ in that the cumulants of the former are $\#P$-hard to compute while the latter are trivial, as they are all zero beyond second order.

\subsection{Photon-number cumulants of Gaussian states in exponential time}
In this section we derive an exponential-time algorithm to calculate photon-number cumulants of Gaussian states. 
A naive evaluation of the cumulants can be obtained by using Eq.~\eqref{eq:cumulant_general} giving cumulants in terms of moments. For $\ell$-order cumulants involving as many modes (and assuming for simplicity zero displacements) this will require to evaluate the partitions of precisely $\ell$ numbers, which are given by the $\ell-$th Bell number, $B_\ell$, which asymptotically scales as (cf. page 108 of De~{B}ruijn~\cite{de1981asymptotic})
\begin{align}
	B_\ell \sim \exp\Bigg[ &\ell \log \ell - \ell\log \log \ell -\ell + \frac{\ell\log \log \ell}{\log \ell}  \Bigg. \\
	& \Bigg. + \frac{\ell}{\log \ell} + \frac{\ell}{2} \left( \frac{\log \log \ell}{\log \ell} \right)^2 + O\left( \frac{\ell\log \log \ell}{ (\log \ell)^2} \right)\Bigg], \nonumber 
\end{align}
which grows exponentially in $\ell$.
Moreover, this requires the evaluation of Hafnians of size as large as $2\ell$, making this method highly impractical.

A second alternative is to use the definition of the Montrealer, which again for $\ell$ modes will require $(2\ell-2)!!$ sums. However, one can do better by writing down the cumulant-generating function (recall Eq.~\eqref{eq:M} and Eq.~\eqref{eq:MM})
\begin{subequations}\label{eq:K}
	\begin{align}
		\mathcal{K}(\bm{t}) &= \log \mathcal{M}(\bm{t})\label{eq:Ma} \\
		&= \tfrac12 \bar{\bm{\zeta}}^\dagger \bm{G} [\I_{2 \ell} - \bm{G} \bm{\Sigma}]^{-1} \bar{\bm{\zeta}} - \tfrac12 \log \det\left( \I_{2 \ell} - \bm{\Sigma} \bm{G} \right) \\
		&= \tfrac12 \sum_{i=0}^\infty \bar{\bm{\zeta}}^\dagger \bm{G} (\bm{G}\bm{\Sigma})^i \bar{\bm{\zeta}} + \tfrac12 \sum_{i=1}^\infty \frac{\text{tr}\left[ \left( \bm{G} \bm{\Sigma} \right)^i\right]}{i} \\
		&= \tfrac12 \sum_{i=0}^\infty \bar{\bm{\zeta}}^\dagger \bm{G} (\bm{G}\X\A)^i \bar{\bm{\zeta}} + \tfrac12 \sum_{i=1}^\infty \frac{\text{tr}\left[ \left( \bm{G} \X \A \right)^i\right]}{i},
	\end{align}
\end{subequations}
where we have used the Taylor series of $(1-x)^{-1}$ and the Mercator identity linking Determinants and power-traces, and recalled that $\bm{X}\S = \A$.

With this equation we can write the Montrealer of any adjacency matrix $\A$ as
\begin{subequations}\label{eq:expmtl}
	\begin{align}
		\mtl \A &= \left. \prod_{i=1}^{\ell} \left(\frac{\partial}{\partial t_i}\right)  \mathcal{K}(\bm{t})  \right|_{\bm{t} = \bm{0}} \label{eq:mtla} \\
		&= \left.  \tfrac12 \prod_{i=1}^{\ell} \left(\frac{\partial}{\partial t_i}\right) \sum_{i=1}^\infty \frac{\text{tr}\left[ \left( \bm{G} \X \A \right)^i\right]}{i} \right|_{\bm{t} = \bm{0}} \label{eq:mtlb}\\
		& =  \left.  \frac{1}{2\ell} \prod_{i=1}^{\ell} \left(\frac{\partial}{\partial t_i}\right)  \text{tr}\left[ \left( \bm{G} \X \A \right)^{\ell}\right]\right|_{\bm{t} = \bm{0}} \label{eq:mtlc}\\
		& =   \left. \prod_{i=1}^{\ell} \left(\frac{\partial}{\partial t_i}\right)  \frac{\text{tr}\left[ \left( \bm{T} \X \A \right)^{\ell}\right]}{2\ell}\right|_{\bm{t} = \bm{0}}, \label{eq:mtld}
	\end{align}
\end{subequations}
where 
\begin{align}\bm{T} = \left[ \oplus_{i=1}^\ell {t_i} \right] \oplus \left[ \oplus_{i=1}^\ell {t_i} \right].
\end{align}
To arrive at Eq.~\eqref{eq:mtlc} we noted that in the sum of Eq.~\eqref{eq:mtlb} only the term with power $i=\ell$ will contribute. To arrive to Eq.~\eqref{eq:mtld} we noted that $e^t - 1 = \sum_{i=1}^\infty \tfrac{t^i}{i!} = t + O(t^2)$.

Having expressed the quantity of interest as a polynomial of degree $\ell$ in the indeterminates $t_1,t_2,\ldots,t_\ell$ we can employ the methods from Ref.~\cite{bax1998finite,bulmer2023master} to write
\begin{align}
	\mtl \A &= \frac{1}{2\ell}\sum_{\bm{i} \in \mathcal{P}([\ell])} (-1)^{|\bm{i}|+\ell} \text{tr}\left[ \bm{\Sigma}[\bm{i}]^\ell \right],
\end{align}
where $[\ell] = {1,2,\ldots,\ell}$, $\mathcal{P}(A)$ is the powerset of $A$, i.e., the set of all subset of $A$ (recall that if $A$ has $|A|$ elements then $\mathcal{P}(A)$ has $2^{|A|}$ elements), $\bm{\Sigma}[\bm{i}]$ is the matrix obtained from $\S = \bm{X} \bm{A}$ by keeping the rows and columns $[i_1,\ldots i _{|\bm{i}|}, i_{1}+\ell, \ldots,  i _{|\bm{i}|} + \ell]$ where $|\bm{i}|$ is the number of elements in $\bm{i}$. Note that the number of summands in the equation above is precisely $2^\ell$ and the calculation of any of them requires at most $O(\ell^3)$ operations, thus the formula above give an exponential-time algorithm for calculating Montrealers and also for calculating Hamiltonian cycles. The speed of our algorithm is comparable to the state of the art methods of Bj\"orklund et al. (Sec. 1.2 of Ref.~\cite{bjorklund2019generalized})  for calculating the number of Hamiltonian cycles. 
The methods described here are implemented using fast just-in-time compiled \texttt{Python} (via \texttt{Numba}~\cite{lam2015numba}) as of version 0.21 of The Walrus~\cite{gupt2019walrus}.

\section{Distribution of cumulants for Gaussian states prepared with Haar-random interferometers}\label{sec:numerics}

\begin{figure}[!t]
	\centering
	\captionsetup{justification = raggedright, singlelinecheck = false}
	\includegraphics[width = 0.5\textwidth]{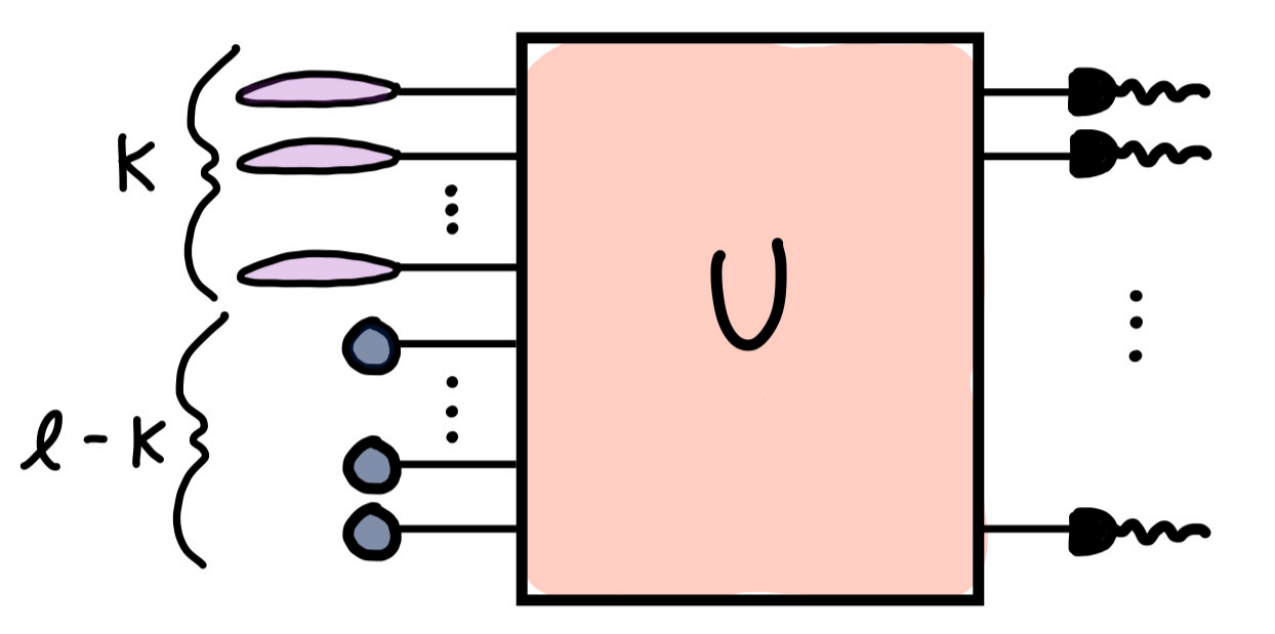}
	\caption{Gaussian boson sampling scheme. $K$ identical single-mode zero displacement Gaussian states are sent through an $\ell$-mode interferometer represented by the unitary $\U$ leaving the remaining $\ell-K$ modes to the vacuum. The output state of the interferometer is sampled using photon-number resolving detectors.}
	\label{fig:GBS_montage}
\end{figure}

We now consider the statistics of cumulants up to the fourth order when $K$ identical single-mode Gaussian states are injected in an $\ell$-mode interferometer as illustrated on Fig.~\ref{fig:GBS_montage}. For the remainder of this article, we will only consider zero displacement inputs although, recent GBS experiments on displaced squeezed states have been reported \cite{thekkadath2022experimental}. 

The input $\ell \times \ell$ phase-sensitive and phase-insensitive matrices, $\M_{\text{in}}$ and $\N_{\text{in}}$ are
\begin{align} \label{eq:mean_eccentricity}
	\N_{\text{in}} = \nb\left(  \I_K \oplus 0_{\ell-K} \right), \quad
	\M_{\text{in}} = \mb \left(\I_K \oplus 0_{\ell-K} \right),
\end{align}
where $\nb = \braket{\ah^\dagger \ah}$ and $\mb = \braket{\ah^2}$ are the expected photon-number and eccentricity of the input modes. 

\begin{figure*}[!t]
	\centering
	\includegraphics[width = \textwidth]{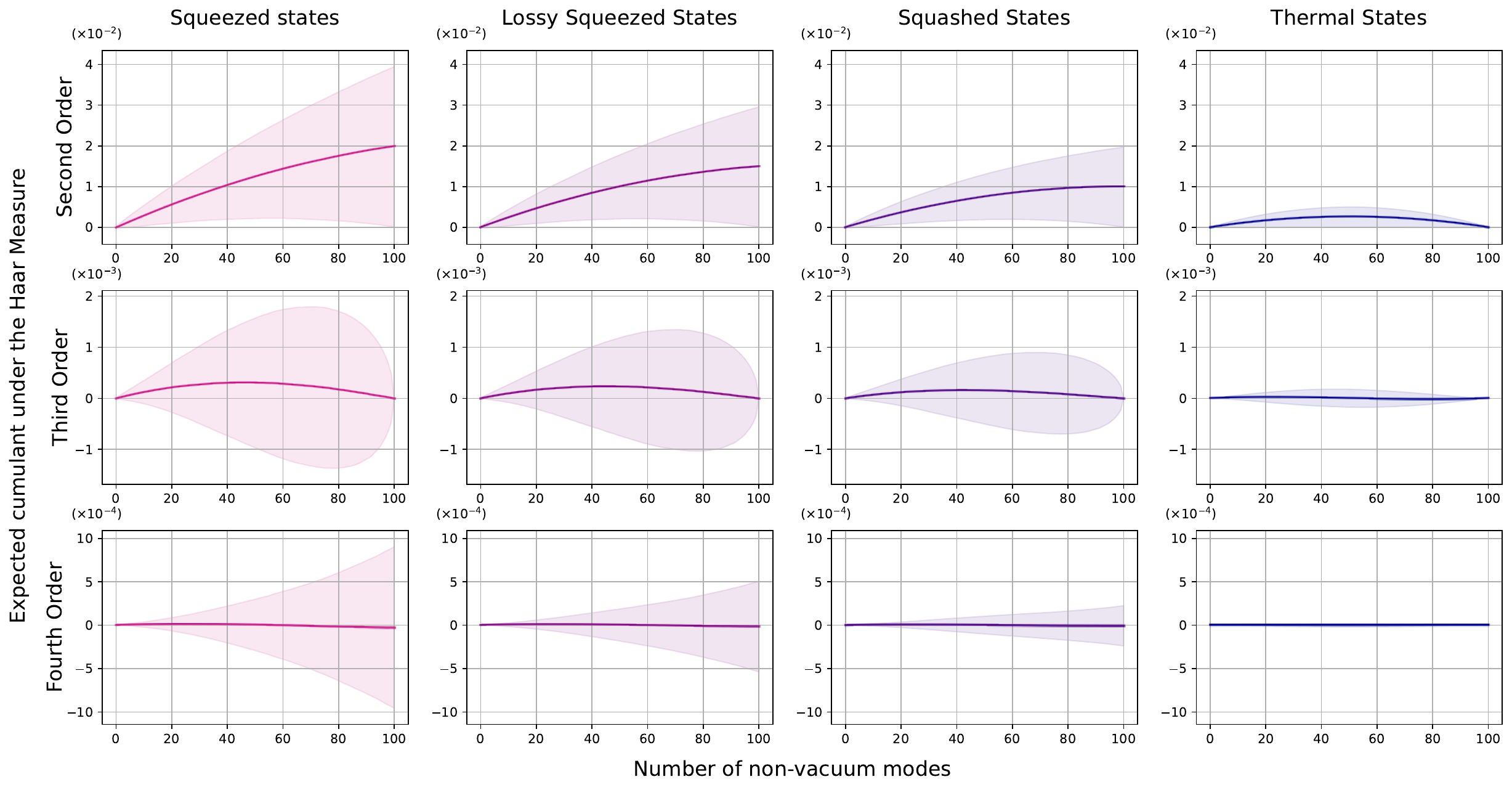}
	\captionsetup{justification = raggedright, singlelinecheck = false}
	\caption{Monte Carlo simulation of the expected cumulant (straight line) and a single standard deviation (shading) over a million unitary matrices $\U$ sampled uniformly at random from the Haar measure for different cumulant orders and different states of light for 100-mode boson samplers using identical single-mode zero displacement states. The mean photon-number in each input state is $\bar{n}=1$  for all studied states.
		For the lossy squeezed state case we consider an input mean photon number in each mode $\bar{n}_{\text{in}} =2$, followed by an energy transmission of $\eta=1/2$ to give a net mean photon number $\bar{n} = \eta  \bar{n}_{\text{in}} = 1$. }
	\label{fig:monte_carlo_cumulant}
\end{figure*}

Single-mode gaussian states can be uniquely specified by their mean photon-number and eccentricity. The relation between these two quantities for all states considered in this section is illustrated in Table \ref{tab:states}. Note that squashed states are classical states for they have non-negative Glauber-Sudarshan P function~\cite{martnez2022classical, qi2020regimes, jahangiri2020point, drummond2014quantum, rahimi2016sufficient, rahimi2015can}. Those classical states exhibit behaviors similar to their close relatives, the squeezed states, which are quantum states, making squashed states a candidate of choice to spoof GBS.
\begin{table}[]
	\begin{tabular}{ccc}
		
		State          & $\mb$ (Eccentricity) & Classical \\ \hline
		Squeezed        & $\sqrt{\nb(\nb+1)}$ & \xmark       \\ 
		Lossy squeezed  & $\sqrt{\nb(\nb+\eta)}$ & \xmark       \\ 
		Squashed        & $\nb$ & \cmark       \\ 
		Thermal         & 0 & \cmark       \\ 
	\end{tabular}
	\captionsetup{justification = raggedright, singlelinecheck = false}
	\caption{Relation between mean photon-number $\nb$ and eccentricity $\mb$ of selected Gaussian states, $\eta$ being the transmission rate of the lossy squeezed state. A state is deemed non-classical if $\mb > \nb$ as it has a negative $P$ function. Note that when $\eta \to 1$ a lossy squeezed state becomes a squeezed state.}
	\label{tab:states}
\end{table}

The transformation law of the phase-insensitive and the phase-sensitive matrices under an interferometer represented by a unitary matrix $\U$ reads
\begin{align}
	\N_{\text{out}} = \U^*\N_{\text{in}}\U^T,\quad 
	\M_{\text{out}} = \U\M_{\text{in}}\U^T.
\end{align}
Interferometers are represented by unitary matrices, thus, it is possible to portray the statistical signature of GBS from many unitaries sampled uniformly at random from the Haar measure which we refer to as Haar-random interferometers. One can, in principle, find analytic expressions of the moments of the cumulants over the Haar measure for arbitrary orders by using Weingarten calculus but we will not do that here for these expressions are unwieldy~\cite{phillips2019benchmarking, puchala2011symbolic, brouwer1996diagrammatic, fukuda2019rtni}. We instead resort to Monte Carlo simulations of the first four orders. This can be done by feeding unitary matrices $\U$ sampled at random from the Haar measure into the cumulant expressions previously found. For each value of $K$, a million such matrices were used to obtain the results in Fig.~\ref{fig:monte_carlo_cumulant}. The shading surrounding the expected cumulant is the simulated standard deviation.
\begin{figure}[!t]
	\centering
	\includegraphics[width = 0.5\textwidth]{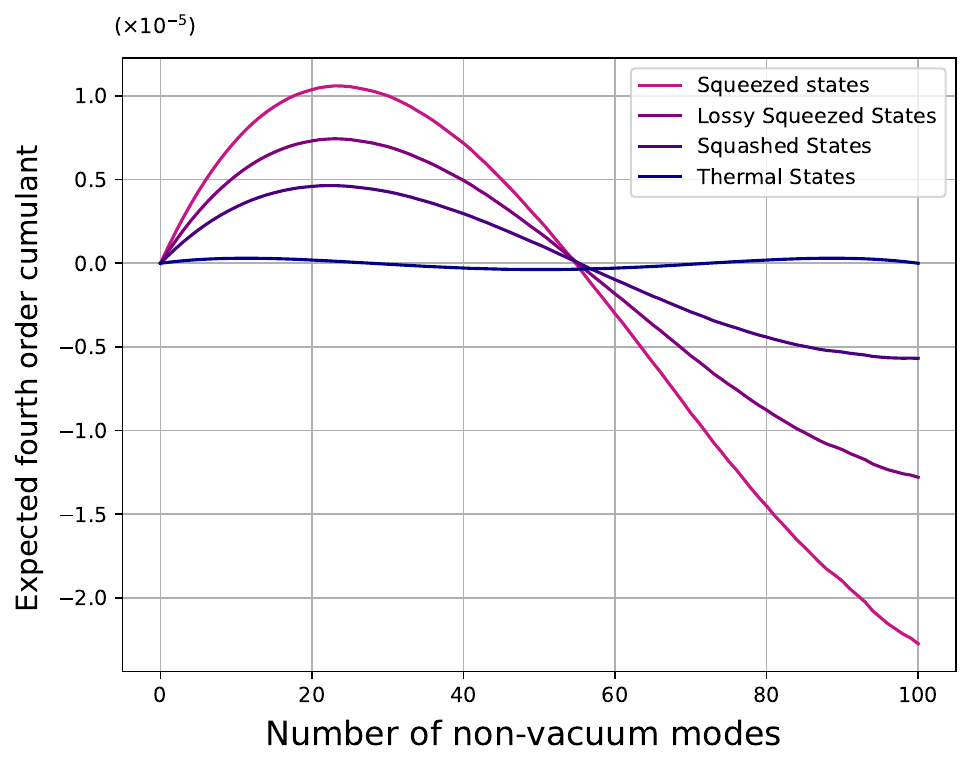}
	\captionsetup{justification = raggedright, singlelinecheck = false}
	\caption{Monte Carlo simulation for 10 million trials of the fourth order cumulant expected value under the Haar measure, $\mathbb{E}_{U}[\langle\langle\hat{n}_1\hat{n}_2\hat{n}_3\hat{n}_4)\rangle\rangle]$, for $0\le K \le \ell =100$. For all shown states, $\bar{n}=1$. It is assumed that there is no displacement in the system.}
	\label{fig:fourth_order}
\end{figure}

We  first note the clear resemblance between squeezed and squashed states, much more so than between squeezed and thermal states. Indeed, as shown in Fig.~\ref{fig:GBS_montage}, both the expected value and the standard deviation of the cumulants, for all considered orders, grow in the same way for squeezed, lossy squeezed and squashed states. The same cannot be said of thermal states, which exhibit less variability than any other state in our simulations. Their expected cumulants are also considerably closer to zero for all values of $K$ than any of the three other states.

The third-order expected cumulants and their standard deviation are in agreement, for $K=\ell$, with our analytical results which predict that odd-order cumulants are precisely zero when all input modes are occupied giving a diagonal $\bm{N}$. For the fourth order cumulants, the standard deviation over the Haar measure is much larger than the expected cumulants. In order to better capture the expected fourth-order cumulants, we compare them directly in Fig.~\ref{fig:fourth_order}. 
As it is seen in this figure, input-states that are phase sensitive behave similarly,  essentially all being a scaled version of each other. This of course is not the case for thermal states, which are not phase-sensitive and thus the behavior of their mean fourth order cumulant is flat relative to the other states.

\section{Conclusion}\label{sec:fini}
In this work, we developed a closed-form expression as well as a graph representation for the moments and cumulants of Gaussian states when measured in the photon-number basis. We found that the moments can be expressed in terms of loop Hafnians or simply in terms of Hafnians for zero displacement Gaussian states. Similarly, we introduced a new matrix function, the Montrealer, to calculate photon-number cumulants of Gaussian states. We show that the calculation of the moments and cumulants expressions are $\#P$-hard. We also provide an exponential time algorithm to calculate the cumulants, and point to recently developed algorithm for Hafnians which allow for a calculation of moments with similar exponential time complexity.

Some interesting properties of the Montrealer were then deduced. For instance, the Montrealer of any adjacency matrix built from an odd-sized diagonal phase-insensitive matrix is zero. Thus, except for the mean, the odd-order cumulants for different modes, for zero-displacement Gaussian states, are all identically zero. Counterintuitively, this implies that a perfect GBS machine built with perfect squeezers in every input of a lossless interferometer will have zero odd-order cumulants, while a noisy machine (with potentially zero entanglement) in which some of the input modes of interferometer are fed with vacuum will have finite non-zero values for these quantities.

We then performed a statistical analysis, through Monte Carlo simulations, comparing the cumulants up to the fourth order for four different types of single-mode Gaussian states fed into a subset of the inputs of an interferometer.  We showed, based on our expression for cumulants, that squashed states are significantly better than thermal states at mimicking photon-number cumulants of lossy and lossless squeezed states, while thermal states perform poorly.

We believe the formulas we derived will be of use to experimentalists aiming at  providing partial evidence for  the correct functioning of a GBS machine, extending what was already done for the experiment in Ref.~\cite{madsen2022quantum}.
Moreover, they give a nice duality relation between probabilities of Gaussian states, given in terms of loop Hafnians of the precision matrix, the inverse of the covariance matrix, and their moments, given in terms of loop Hafnians of the covariance matrix.

\section*{Acknowledgements}
We thank J. Mart\'inez-Cifuentes and M. Doyon for help in preparing the figures. We also thank J. Mart\'inez-Cifuentes and K. Chinni for providing feedback on the manuscript and E. Fitzke for providing the code to calculate moments using automatic differentiation. The authors thank the Minist\`ere de l’\'Economie et de l’Innovation du Qu\'ebec and the Natural Sciences and Engineering Research Council of Canada for their financial support.
N.Q. thanks S. Duque Mesa, J.F.F. Bulmer and D. Grier for insightful discussions.
Y.C. thanks the Fonds de recherche du Qu\'ebec and Polytechnique Montr\'eal's Unit\'e de participation et d'initiation \`a la recherche program for their financial support. 

\appendix

\section{Quadrature and ladder operators}\label{app:quadratures}
The hermitian quadrature operators and annihilation and creation operators are related as follows
\begin{align}
	\ah_j = \frac{\qh_j+i\ph_j}{\sqrt{2\hbar}}\,,\quad
	\ch_j = \frac{\qh_j-i\ph_j}{\sqrt{2\hbar}}\,,
\end{align}
where $\hbar$ is a convenient scale parameter for the variances of the vacuum state.

The quadrature operators can also be collected into a single $2\ell$-dimensional vector, that is,
\begin{align}
	\rh = \left(\qh_1,\dots, \qh\l, \ph_1, \dots, \ph\l\right)^T.
\end{align}
which is related to the vector of creation and annihilation operators as follows
\begin{align} \label{eq:relation_opeartor}
	\zh = \frac{1}{\sqrt{\hbar}}\R\rh,
\end{align}
where we introduced the unitary matrix
\begin{align}
	\bm{R} = \frac{1}{\sqrt{2}} \begin{pmatrix}
		\I\l & i \I\l \\
		\I\l & -i \I\l
	\end{pmatrix}.
\end{align}
We can write the commutation relations of the quadrature operators succinctly as
\begin{align}
	[\hat{r}_i, \hat{r}_j] = i \hbar \Omega_{ij} \text{ with } \O = \begin{pmatrix}
		0_\ell & \I\l    \\
		- \I\l & 0_\ell  
	\end{pmatrix}.
\end{align}
The relation between the two representations' covariance matrices, $\V$ and $\S$, for the quadratures representation and the annihilation and creation representation respectively, is
\begin{align}
	\S^{(s)}= \frac{1}{\hbar}\R\V^{(s)}\R\d.
\end{align}
Note that the covariance matrices obey their respective uncertainty relation (cf. Sec. 3.4. of Serafini~\cite{serafini2017quantum})
\begin{align}
	\S^{(s)} + \tfrac12 \Z + \tfrac{s}{2} \I_{2 \ell} &\geq 0,\\
	\V^{(s)} + i \tfrac{\hbar}{2} \bm{\Omega} + \tfrac{s \hbar }{2} \I_{2 \ell} &\geq 0.
\end{align}
An immediate consequence of these inequalities is that
\begin{align}
	|M_{ij}| \leq \min\left\{\sqrt{N_{ii} (1+N_{jj})}, \sqrt{N_{jj} (1+N_{ii})} \right\},
\end{align}
and moreover if for some $i$, $N_{ii} = 0$ then $M_{ij}=0$
$\forall j$.

	\bibliographystyle{unsrtnat}

\bibliography{bib.bib}

\end{document}